\documentclass{article}

\usepackage[nonatbib]{neurips_2024}

\usepackage[square,numbers]{natbib}
\usepackage[utf8]{inputenc} 
\usepackage[T1]{fontenc}    
\usepackage{hyperref}       
\usepackage{url}            
\usepackage{booktabs}       
\usepackage{amsfonts}       
\usepackage{nicefrac}       
\usepackage{microtype}      
\usepackage{fontawesome5}
\usepackage{graphicx}
\usepackage{amsmath}
\usepackage{bm} 
\usepackage{multirow}
\usepackage{makecell}
\usepackage{enumitem}
\usepackage[table,xcdraw]{xcolor}
\definecolor{table_gray}{HTML}{EFEFEF}
\usepackage{float}
\usepackage[labelformat=simple]{caption}
\usepackage{arydshln}
\usepackage{lineno}
\usepackage{svg}
\usepackage{upgreek}
\usepackage{multicol}

\DeclareCaptionLabelFormat{customlabel}{\textbf{#1 #2}}
\newcommand{\inst}[1]{\textsuperscript{#1}}

\title{Mamba-based Deep Learning Approach for Sleep Staging on a Wireless Multimodal Wearable System without Electroencephalography}

\author{
    Andrew H. Zhang\normalfont$^\dagger$\inst{ 1}\inst{ 2}\inst{ 3}\textsuperscript{ \faEnvelope[regular]}
    \And 
    Alex He-Mo\normalfont$^\dagger$\inst{ 1}\inst{ 2}\textsuperscript{ \faEnvelope[regular]}
    \And 
    Richard Fei Yin\normalfont$^\dagger$\inst{ 1}\inst{ 2}\textsuperscript{ \faEnvelope[regular]}
    \And \quad
    Chunlin Li\normalfont\inst{ 1}
    \And 
    Yuzhi Tang\normalfont\inst{ 1}
    \And 
    Dharmendra Gurve\normalfont\inst{ 2}
    \And 
    Veronique van der Horst\normalfont\inst{ 5}
    \And
    Aron S. Buchman\normalfont\inst{ 6}
    \And    
    Nasim Montazeri Ghahjaverestan\normalfont\inst{ 2}\inst{ 7}
    \And
    Maged Goubran\normalfont\inst{ 1b}\inst{ 2b}
    \And 
    Bo Wang\normalfont\inst{ 1bc}\inst{ 3}\inst{ 4}
    \And
    Andrew S. P. Lim\normalfont\inst{ 1a}\inst{ 2a}\textsuperscript{ \faEnvelope[regular]}
}

\begin{document}
\nolinenumbers
\maketitle
\begin{center}
{\small
    {\inst{1}University of Toronto Dept. of \inst{a}Medicine, \inst{b}Medical Biophysics, \inst{c}Computer Science\hfill Toronto, ON, Canada
    \\
    \inst{2}Sunnybrook Research Institute \inst{a}Dept. of Medicine Neurology Div., \inst{b}Physical Sciences\hfill Toronto, ON, Canada
    \\ 
    \inst{3}Vector Institute for Artificial Intelligence \hfill Toronto, ON, Canada
    \\
    \inst{4}University Health Network \hfill Toronto, ON, Canada
    \\ 
    \inst{5}Beth Israel Deaconess Medical Center \hfill Boston, MA, USA
    \\ 
    \inst{6}Rush University Medical Center \hfill Chicago, IL, USA
    \\ 
    \inst{7}Queen's University Dept. of Electrical \& Computer Engineering \hfill Kingston, ON, Canada}\\
}

\vspace{3mm}

\textsuperscript{ \faEnvelope[regular]} \texttt{\{andrewhz.zhang, alex.hemo, r.yin\}@mail.utoronto.ca}\\
\texttt{andrew.lim@utoronto.ca}
\end{center}
\vfill
\def\thefootnote{$\dagger$}\footnotetext{These authors contributed equally.}

\begin{abstract}
\vspace{-4mm}
  \textbf{Study Objectives}\\
 We investigate a Mamba-based deep learning approach for sleep staging on signals from ANNE One (Sibel Health, Evanston, IL), a non-intrusive dual-module wireless wearable system measuring chest electrocardiography (ECG), triaxial accelerometry, and chest temperature, and finger photoplethysmography and finger temperature.\vspace{0.5em}
  
  \textbf{Methods}\\
We obtained wearable sensor recordings from 357 adults undergoing concurrent polysomnography (PSG) at a tertiary care sleep lab. Each PSG recording was manually scored and these annotations served as ground truth labels for training and evaluation of our models. PSG and wearable sensor data were automatically aligned using their ECG channels with manual confirmation by visual inspection. We trained a Mamba-based recurrent neural network architecture on these recordings. Ensembling of model variants with similar architectures was performed.\vspace{0.5em}

  \textbf{Results}\\
  After ensembling, the model attains a 3-class (wake, non rapid eye movement [NREM] sleep, rapid eye movement [REM] sleep) balanced accuracy of 84.02\%, F1 score of 84.23\%, Cohen's $\kappa$ of 72.89\%, and a Matthews correlation coefficient (MCC) score of 73.00\%; a 4-class (wake, light NREM [N1/N2], deep NREM [N3], REM) balanced accuracy of 75.30\%, F1 score of 74.10\%, Cohen's $\kappa$ of 61.51\%, and MCC score of 61.95\%; a 5-class (wake, N1, N2, N3, REM) balanced accuracy of 65.11\%, F1 score of 66.15\%, Cohen's $\kappa$ of 53.23\%, MCC score of 54.38\%.\vspace{0.5em}
  
  \textbf{Conclusions}\\
  Our Mamba-based deep learning model can successfully infer major sleep stages from the ANNE One, a wearable system without electroencephalography (EEG), and can be applied to data from adults attending a tertiary care sleep clinic.
\end{abstract}

\vspace{-5mm}

\begin{center}
    \large\textbf{Keywords:}\quad Deep Learning\quad Sleep Staging\quad Wearable Devices
\end{center}

\pagebreak
\section*{Statement of Significance}
Sleep is traditionally measured by in-lab polysomnography followed by visual identification of sleep stages by specialized technologists.  However, the need for specialized personnel and equipment makes polysomnography difficult to scale, and poses a barrier to many older adults.  To address this, we developed and evaluated in 357 patients attending a tertiary care sleep clinic a Mamba-based AI model for inferring sleep stage from the ANNE One sensor system -- a pair of wearable minimally intrusive sensors placed on the chest and finger.  We show good performance at distinguishing sleep stages across a range of ages and co-morbid sleep pathologies, suggesting that this may represent a scalable, accurate, minimally intrusive approach to the ambulatory assessment of sleep in clinical populations.
\vspace{4em}
\section*{Graphical Abstract}
\centerline{\includegraphics[width=1.1\textwidth]{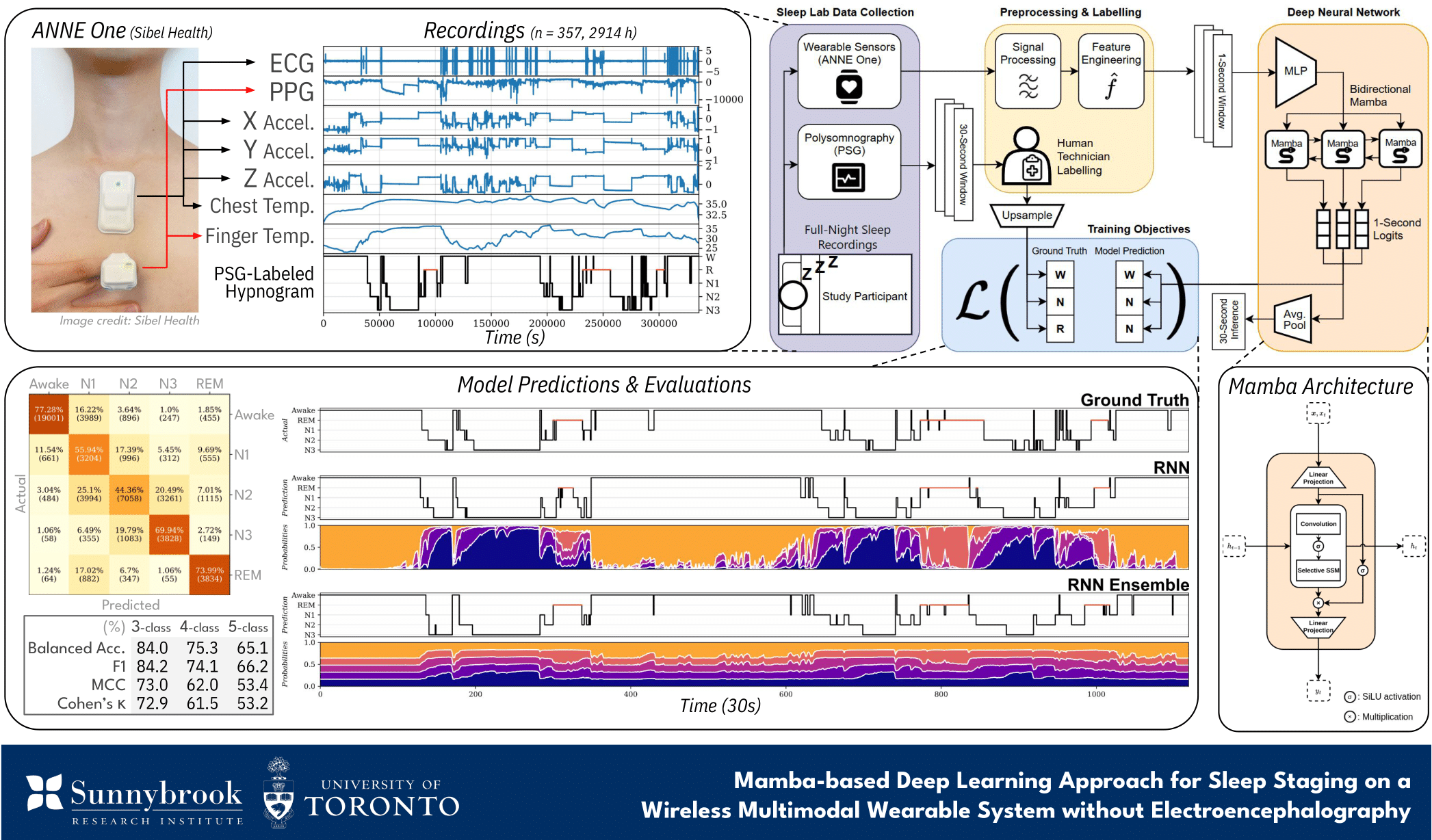}}
\thispagestyle{empty}
\pagebreak

\section{Introduction}
\subsection{Sleep Staging and Polysomnography}
Sleep  is essential to physical and mental health. Quantitative evaluation of sleep architecture and physiology is central to the diagnosis of many sleep disorders, playing an important role in evaluating individuals with and at risk for neurodegenerative diseases \cite{iranzo2015sleep}. Sleep is traditionally and most reliably measured by polysomnography (PSG), in which brain activity, eye movements, heart rate (HR), muscle tone, and breathing patterns are recorded simultaneously by sensors attached to a sleeping patient in a laboratory environment. This is followed by sleep staging, the categorization of time periods into sleep stages based on a trained sleep technician's visual inspection of the collected physiological data. The exact sleep staging methodology depends on the specific framework; under the American Academy of Sleep Medicine's (AASM) \cite{aasm} conventions, recordings are divided into 30-second windows, or "epochs". Electroencephalography (EEG), electromyography (EMG), and electro-oculography (EOG) are used to classify epochs into three stages: awake, rapid eye movement (REM) sleep, and non-REM (NREM) sleep, which can be further subcategorized as N1, N2, and N3 sleep, with N1 as the lightest and N3 as the deepest NREM sleep. EEG is the principal signal used for sleep staging \cite{aasm, rechtschaffen1968brain, imtiaz2021systematic}.

\subsection{Wearable Devices and Automated Sleep Staging}

The need for specialized personnel, equipment, and laboratory space makes PSGs expensive to conduct and difficult to scale. The cumbersome sensor setup and unfamiliar resting environment may also discomfort patients, resulting in sleep that is unrepresentative of that in the home environment. Particularly sick or frail older adults may be physically unable to attend in-laboratory PSG, in addition to those living in remote areas without PSG services. Furthermore, monitoring patients throughout the night and manually labelling sleep epochs is a time-inefficient process that limits the volume of sleep analyses.

In response to these challenges, wearable devices have been developed that record a subset of conventional PSG signals. Devices that record EEG have the advantage of directly measuring brain activity but the disadvantage of the discomfort associated with sensors on the head similar to PSGs.  On the other hand, EEG-less devices are less obtrusive and more easily set up for in-home use without a technician, but infer rather than directly measure sleep stage.  

There has been considerable interest in pairing wearable devices with automated sleep stage inference in order to minimize the need for specialized technologists for labeling and hence improve scalability.  Indeed, even for the analysis of conventional in-laboratory PSG, automated sleep staging has been explored as a means of addressing the time-intensiveness of sleep epoch labelling. The focus on pattern recognition and loosely rule-based nature of sleep staging is ideal for machine learning approaches which have employed model architectures like convolutional neural networks (CNNs) \cite{tsinalis2016automatic}, recurrent neural networks (RNNs) \cite{malafeev2018automatic}, hybrids of the previous two (e.g. convolutional recurrent neural networks (CRNN)) \cite{seo2020intra}, transformers \cite{phan2022sleeptransformer}, and graph neural networks \cite{jia2020graphsleepnet}. Deep learning models trained on PSG data, most notably the EEG signal or some combination of EEG, EMG, and EOG, have surpassed inter-scorer agreement levels, with 5-class accuracy and Cohen's $\kappa$ as high as 92\% and 86\% respectively, and can make predictions orders of magnitude faster than a human evaluator at the cost of sacrificed interpretability \cite{fiorillo2019automated, phan2022automatic}.


\subsection{Recurrent Neural Networks and Mamba}
RNNs are a class of neural networks that takes in a sequence of vectors (e.g. a time series of pulse waves or breathing) and produce a vector output per input, allowing them to execute a wide breadth of time-series tasks. Prevalent modern RNN architectures include long short-term memory (LSTM) \cite{hochreiter1997long} and gated recurrent units (GRU) \cite{cho2014learning}, which are known to forget very long-range dependencies and whose architecture cannot exploit GPU acceleration, limiting the training speed. Transformers \cite{vaswani2017attention} address these problems but scale poorly in speed for longer sequences both for training and inference.

Mamba \cite{mamba} as a new generation of RNN provides a solution to these issues as a type of state space model (SSM) \cite{gu2021combining} that combines the excellent long-range dependency modeling of SSMs with a time-dependent selection mechanism that allows for focusing on particular features, as well as a hardware-friendly algorithm that is computationally efficient (see \ref{sec:mamba} for details). Empirically, Mamba has achieved state-of-the-art performance rivaling that of transformers on supervised and unsupervised metrics from datasets of text, audio waveforms, and DNA sequences \cite{mamba}, while Mamba-based models have been adopted to perform sleep staging using EEG and other PSG signals to high degrees of accuracy \cite{zhang2024mssc, zhou2024bit}.


\begin{figure}[H]
\centerline{\includegraphics[width=0.9\textwidth]{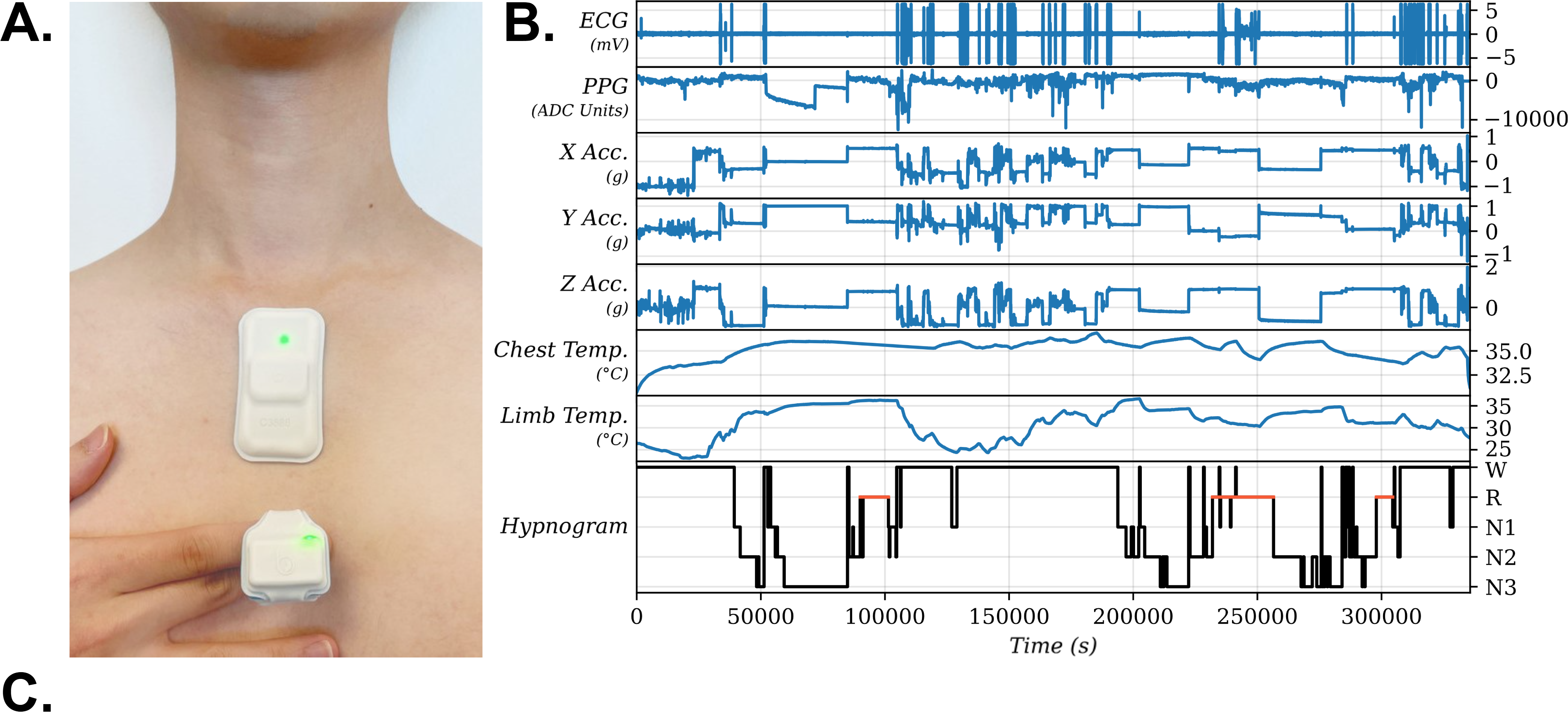}}
\end{figure}

\vspace{-3.5em}
\begin{table}[H]
    \begin{center}
    \footnotesize
    \begin{tabular}{lrrc}
         \rowcolor[HTML]{EFEFEF}
         \Xhline{3\arrayrulewidth}
         \textit{Statistic}\phantom{\Large{A}} & $\mu$ & $\sigma$ & \textit{Range / Count} \\
         \hline

         Recordings \phantom{\Large{A}} & & & 357 \\

         Male/Female & & & 181/176 (51\%/49\%) \\
         
         Recording Duration [h] & 8.16 & 1.39 & 1.75 $-$ 12.94 \\
         
         Awake [\%] & 40.96 & 16.58 & 10.47 $-$ 95.24 \\
         
         N1 Sleep [\%] & 11.13 & 6.24 & 1.05 $-$ 38.42 \\
         
         N2 Sleep [\%] & 28.37 & 10.61 & 0.58 $-$ 63.59 \\
         
         N3 Sleep [\%] & 10.85 & 8.12 & 0.00 $-$ 54.83 \\
         
         REM Sleep [\%] & 8.70 & 6.08 & 0.00 $-$ 29.14 \\
         
         Age [y] & 57.49& 17.56 & 18.17 $-$ 92.07 \\
         
         Apnea-Hyponea Index (AHI) & 11.54 & 17.21 & 0.00 $-$ 127.00 \\
         
         Total Sleep Time (TST) [h] & 5.10 & 1.46 & 0.66 $-$ 9.16 \\
         
         Body Mass Index (BMI) [kg $\text{m}^{-2}$] & 27.88 & 5.59 & 16.90 $-$ 50.40 \\
         
         Periodic Limb Movement Index (PLMI) & 15.01 & 26.99 & 0.00 $-$ 169.50 \\

         Sleep Efficiency [\%] & 71.83 & 17.91 & 11.60 $-$ 98.50 \\
         
         \Xhline{3\arrayrulewidth}
    \end{tabular}
    \end{center}
\end{table}

\vspace{-2.5em}
\begin{figure}[H]
\caption{\textbf{A.} The finger and chest modules of ANNE One (Image by Sibel Health \cite{sibel}). \textbf{B.} Raw physiological signals recorded by ANNE One during a full night sleep recording and its accompanying hypnogram. \textbf{C.} Demographics and sleep characteristics of the subjects used in this study.}
  \label{fig:anne}
\end{figure}
\vspace{-2.2em}

\subsection{Contribution}

This paper applies a Mamba-based deep learning approach to the problem of sleep staging using data from the ANNE One \cite{sibel} sensor. ANNE One, as seen in Figure \ref{fig:anne}A, is a flexible, minimally-intrusive, and clinical-grade wireless dual-module wearable system by Sibel Health. It is composed of a finger module, which measures photoplethysmography (PPG) and limb temperature, and a separate module attached to the chest via an adhesive, which measures electrocardiography (ECG), triaxial accelerometry, and chest temperature. The raw signals of a typical recording are plotted in Figure \ref{fig:anne}B alongside a hypnogram, a technician-labelled plot of sleep stage over time.

A prior study centered on ANNE One \cite{anne_personal} exists, which primarily emphasized the development of person-specific models which were trained and evaluated on data from the same patient; in contrast, the population model, trained and evaluated on data from different patients, showed much poorer performance. We aim to develop a superior population model to achieve accurate sleep staging from ANNE One signals recorded from previously unseen patients.

There is a considerable body of work applying machine learning to the problem of sleep stage inference from non-EEG signals. For the most part, these have been trained and tested using the cardiopulmonary sensors of conventional PSG \cite{radha2019sleep, sun, sleepppg, pini, topalidis2023pulses, jones2024expert, kazemi2024improved}, which are typically artifact-free.  Relatively few studies have applied these approaches to data from wearable devices, which are in general more prone to poor signal quality and signal loss. Consequently, such studies often need to reject epochs of poorer-quality signal prior to model training and prediction.  Moreover, these works have focused on analysis of data from single sensors, typically ECG or PPG \cite{silva, zhang2018sleep}.   Finally, models developed for wearable sensors have largely trained on data from relatively young and healthy volunteers, rather than patients presenting to sleep clinics, raising a question of generalizability to clinical populations.  

The current study builds on and extends this body of work in several ways.  First, we leverage the ANNE One devices' multi-sensor nature, which may in principle allow for predictions even in settings where one or more sensors are reading low-quality signals.  Second, we train and test on data from relatively older patients with a range of sleep disorders  (41\% were over the age of 65 and 46\% had sleep apnea; Figure \ref{fig:anne}C) rather than younger and healthier patients, ensuring applicability to older adults with sleep complaints.  This is particularly important as age, sleep, and neurological conditions can all plausibly alter the relationship between sleep stage and autonomic, cardiac, and pulmonary function, meaning that models trained on younger adults may generalize poorly to older adults. Finally, this work broadens the Mamba architecture to EEG-less sleep staging by exploring whether Mamba's relative performance increases \cite{zhang2024mssc, zhou2024bit} are also noticed in this context.




\section{Method}

\begin{figure}[H]
  \centering
  \includegraphics[width=\textwidth]{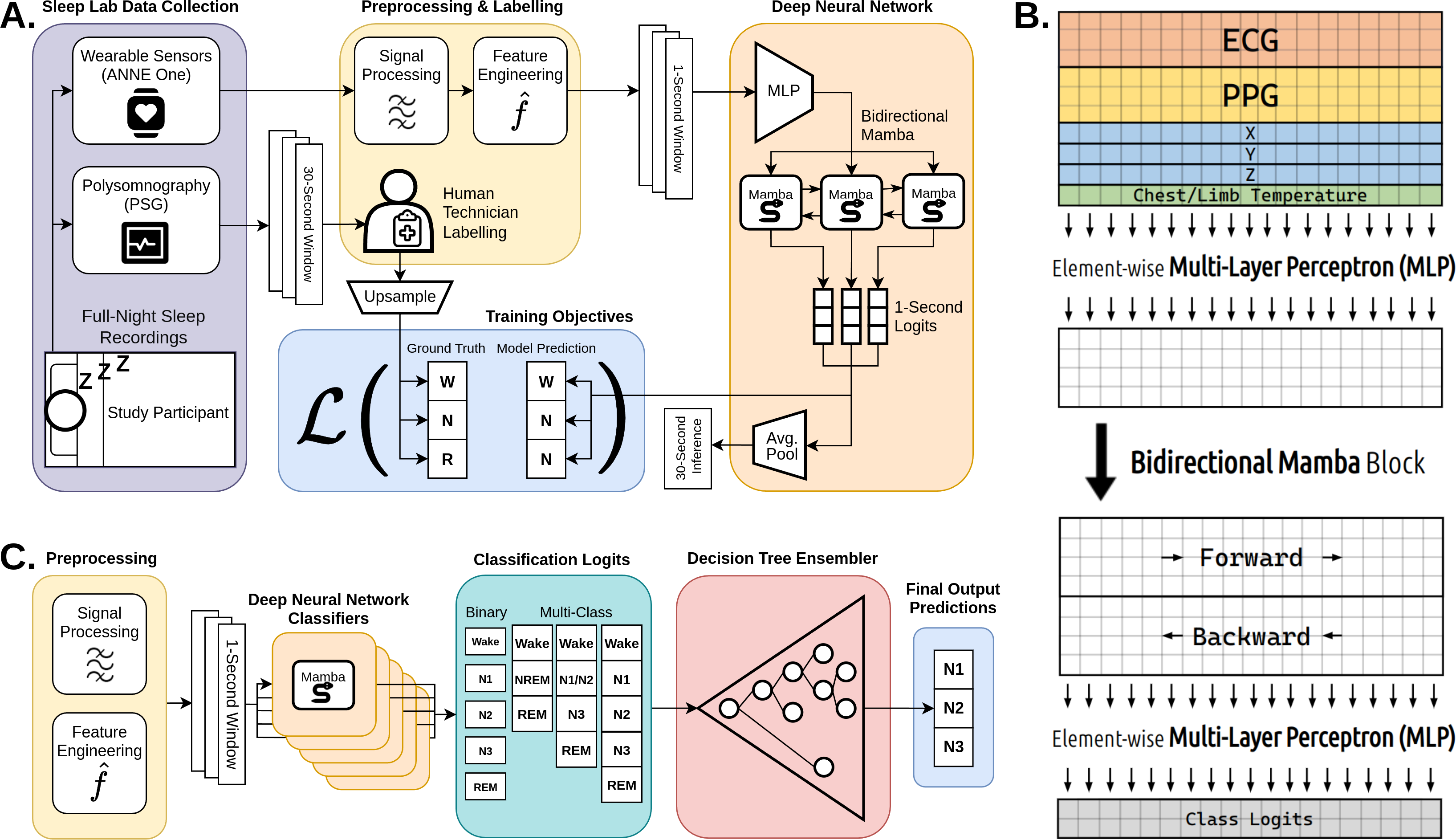}
  \caption{\textbf{A.} Overview of deep learning pipeline of this study. \textbf{B.} Implementation of the RNN architecture in this study. Batch normalizations precede each of the five linear layers in the MLPs. LeakyReLU is the activation function (see \ref{sec:detailed-architecture} for details). \textbf{C.} The inference-time ensembling pipeline.}
  \label{fig:pipelines}
\end{figure}

\subsection{Dataset}



ANNE wearable sensors were used to collect full-night recordings from 360 adults undergoing concurrent clinical PSG at a tertiary care clinical sleep laboratory in Sunnybrook Health Sciences Centre, Toronto, Canada. PSGs were recorded according to AASM practice parameters for in-labratory PSGs with a Grael PSG system (Compumedics, Victoria, Australia).  The study was conducted in accordance with the latest version of the Declaration of Helsinki. All subjects provided written informed consent for the full study protocol. 

Participants were first instrumented with the ANNE One, whose recordings were started ahead of the set-up and initiation of PSG recordings. In order to mimic what would occur in a home environment, the ANNE signals were not monitored in real time, and no attempt was made to correct sensor-related issues (e.g. poor skin contact) overnight. Upon completion, each PSG recording was manually labeled by one of four technologists according to AASM standards \cite{aasm}. In addition, the portion of the ANNE recordings preceding the onset of PSG recording was labeled as wake based on direct observation of participants by study staff. PSG and wearable sensor data were automatically aligned by their shared ECG signals using cross-correlation analysis and verified by visual inspection.

Three participants who did not sleep at all were excluded from analysis, as were time periods prior to the sensor placement or after the sensors were taken off, leaving a total of 357 patients with 2914 hours of recording.  The characteristics of the study's participants are outlined in Figure \ref{fig:anne}C. Generally, older participants (age $\geq$ 65) and those with sleep apnea had poorer sleep quality, with lower total sleep time, poorer sleep efficiency, and increased N1 sleep time relative to other sleep stages (Figure \ref{sec:summary-statistics-detailed})

The machine learning task is defined as multiclass 1D-signal segmentation, wherein the ANNE One signals constitute the features, the PSG-annotated sleep stages form the labels, and class predictions are made per point in time to form a hypnogram as demonstrated in Figure \ref{fig:anne}B.

\subsection{Preprocessing}



The ECG data were recorded at 512 Hz and processed by a multilevel 1D Coiflet transform with 5 moments. Peaks in the resulting signal were identified and used to define windows; within each window, the original ECG signal was searched for local maxima and minima to determine the precise timing of the R-peak. The wavelet-transformed ECG data were subsequently downsampled to 100 Hz for additional analyses. We then computed an ECG signal quality index (SQI) as follows.  Trains of 5 or more physiological QRS complexes were identified on the basis of relatively stable HR (differing by no more than 10 beats per minute between consecutive beats) and amplitude (standard deviation of amplitude $<25$).  All identified ECG peaks in the recording were then clustered using HDBSCAN \cite{mcinnes2017hdbscan} based on amplitude and morphology; peaks that clustered with the template QRS complexes were considered to have passed quality control (QC).  Segments of ECG of 2 seconds on either side of these peaks were assigned an SQI of 1 while all other segments were given an SQI of 0.  

The PPG data were recorded at 128 Hz, then band-pass filtered between $0.5 - 5 \text{ Hz}$ and downsampled to 100 Hz.  Trains of 5 or more physiological pulse waves were identified on the basis of relatively stable HR (differing by no more than 10 beats per minute between consecutive beats) and amplitude (differing by no more than 2-fold between beats).  These high-quality pulse waves were normalized to a standard amplitude and duration, then clustered using HDBSCAN. For each cluster, the mean of its normalized pulse waves was taken as a template wave for that cluster. Beat-to-beat template matching was then performed to determine the extent to which each pulse wave in the rest of the recording matched one of the template waves, with the lowest sum-squared similarity to the template constituting a measure of signal quality for each PPG pulse wave.  Pulse waves with minimum sum squared $<0.4$ were considered to have passed QC. Segments of PPG signal were set to 1 if they were within 2 seconds of a pulse wave passing QC, and 0 otherwise.

The chest XYZ accelerometry data were recorded at 210 Hz, downsampled to 100 Hz similar to ECG, and band-pass filtered over $0.005-0.2$ Hz with a second-order Butterworth filter.

Log-magnitude spectrograms were taken of each signal by means of a short-time Fourier transform (STFT) with a Hamming window of size equal to the signals' sampling rate (i.e. 100) and with no overlap, creating samples at $k = 1$ Hz (an arbitrary value that was determined experimentally). Only the first 12 frequency bins of the XYZ data were kept as the other higher-frequency bins were made redundant by the band-pass filter. Finally, the frequency bins were concatenated with limb/chest temperature, which was recorded at 1 Hz, and sleep stage label, which was created at $\frac{1}{30}$ Hz (as labels are for 30-second windows) and upsampled to 1 Hz by repeating each label 30 times.

\subsection{Model}

A Mamba-based RNN model (summarized in Figure \ref{fig:pipelines}B) was designed to process batches of 2D tensors with axes for the number of $\frac{1}{k}$-second windows and number of features, which include chest temperature, limb temperature, and frequency bins for ECG, PPG, and XYZ. Logits for each sleep stage are returned at each time step. A model with $c$ output classes has $\approx$ 70K parameters which constitute:


\begin{enumerate}[label=\arabic*)]
\item An element-wise multi-layer perceptron (MLP) with $h = 100$ hidden dimensions that diminishes the feature-size axis to $d = 10$ dimensions, where $h$ and $d$ are experimentally determined.
\item A bidirectional three-layer Mamba block. This consists of two Mamba blocks \cite{mambapy} that process time windows sequentially in opposite directions, whose outputs are concatenated in the feature-length axis to allow for a widened receptive field that detects patterns in signals more effectively. 
\item An element-wise MLP that downsizes the feature size dimension into $c$ logits.

\end{enumerate}

In accordance with AASM practice, we form 30-second window predictions by grouping 1-second logits into contiguous buckets of 30, taking the mean logit for each class, and outputting the sleep stage with the largest corresponding logit.

\subsection{Training}
The training, validation, and testing sets were randomly split in a $75\%/10\%/15\%$ ratio to sizes $267/36/54$. During the training process, recordings were clipped to random segments of size 5000 to allow for mini-batch training ($n = 4$) and to prevent the model from overfitting to the high-level structure of most recordings, i.e., wake periods at the start and end with sleep in the middle. The objective function was cross-entropy loss, weighted on inverse class frequency to adjust for the moderate class imbalance shown in Figure \ref{fig:anne}C, then averaged across batches and time windows. The model was optimized with Adam \cite{kingma}, a widely-used adaptive learning rate optimizer, with learning rate 0.001. The parameters from the training iteration that maximized balanced validation accuracy were chosen for the final model.

 

Three models of varying sleep stage resolution were built: a 3-class, 4-class, and 5-class sleep staging model. The 5-class model has classes for wake, N1, N2, N3, and REM sleep, while the 3-class model groups N1, N2, and N3 sleep into non-REM sleep. The 4-class model designates N1 and N2 sleep as "light sleep" as opposed to N3 or "deep sleep", a common practice in automatic sleep staging \cite{imtiaz2021systematic}. Separate models were trained for each class count as to treat each class equally.

\subsection{Ensembling}
For each of the $n$ classes inside a $n$-class ensemble model, a new binary model of the same architecture was trained from scratch to predict the presence or absence of that class. The logits of these binary models were then stacked with logits from the regular 3/4/5-class models. This collection of logits formed the inputs to a suite of decision tree-based architectures (see \ref{sec:detailed-architecture} for details) including AdaBoost \cite{freund1995desicion, scikit-learn}, random forests \cite{ho1998random, scikit-learn}, extra-tree classifiers \cite{geurts2006extremely, scikit-learn}, histogram gradient boosting \cite{cui2021gbht, scikit-learn}, and XGBoost \cite{chen2016xgboost}. This architecture type processes time windows in a time-independent manner, progressively dividing them into smaller and more homogeneous groups by thresholding based on the value of their features (i.e. a logit value for a sleep stage); the thresholds and choice of features are learned parameters. Sleep stage predictions in a single decision tree are made from the plurality sleep label of training set time windows inside a terminal node, while the proportions of each sleep label become the class probabilities. The model among the suite that maximized balanced validation accuracy was selected as the final ensembler model. A visual representation of the ensembling pipeline is presented in Figure \ref{fig:pipelines}C.

\subsection{Statistical Analyses}

Results were compiled of 3-class, 4-class, and 5-class sleep staging for the regular and ensemble RNN models. Overall performance was evaluated by aggregating all test set recordings and computing the balanced accuracy, weighted precision, weighted recall, weighted F1, Cohen's kappa ($\kappa$), and Matthews correlation coefficient (MCC).

For the best model, performance was plotted on a healthy subset of the test set. Ablation studies were carried out using various subsets of the ANNE One sensor data. We also retrained the model without downsampling the data, and considered alternative architectures including use of LSTM instead of Mamba, and consideration of a CRNN model utilizing the same preprocessing steps as our RNN model. For the best 5-class model, Bland-Altman plots were created to evaluate class-wise biases, followed by analyses of per-recording metrics individually as well as across differing signal quality scenarios, sleep disorders, and clinical variables including age, sex, PLMI, AHI, BMI, TST, and sleep efficiency, and sleep disorders including insomnia, restless legs syndrome, and REM sleep behaviour disorder. Because individual recordings did not necessarily contain instances of all five sleep stages, balanced accuracy may not have been a sensible metric for recording-wise comparisons; hence, accuracy was presented in lieu of balanced accuracy for those cases.

\section{Results}

\subsection{Model Evaluation}

\begin{figure}[H]
  \centerline{\includegraphics[width=1.15\textwidth]{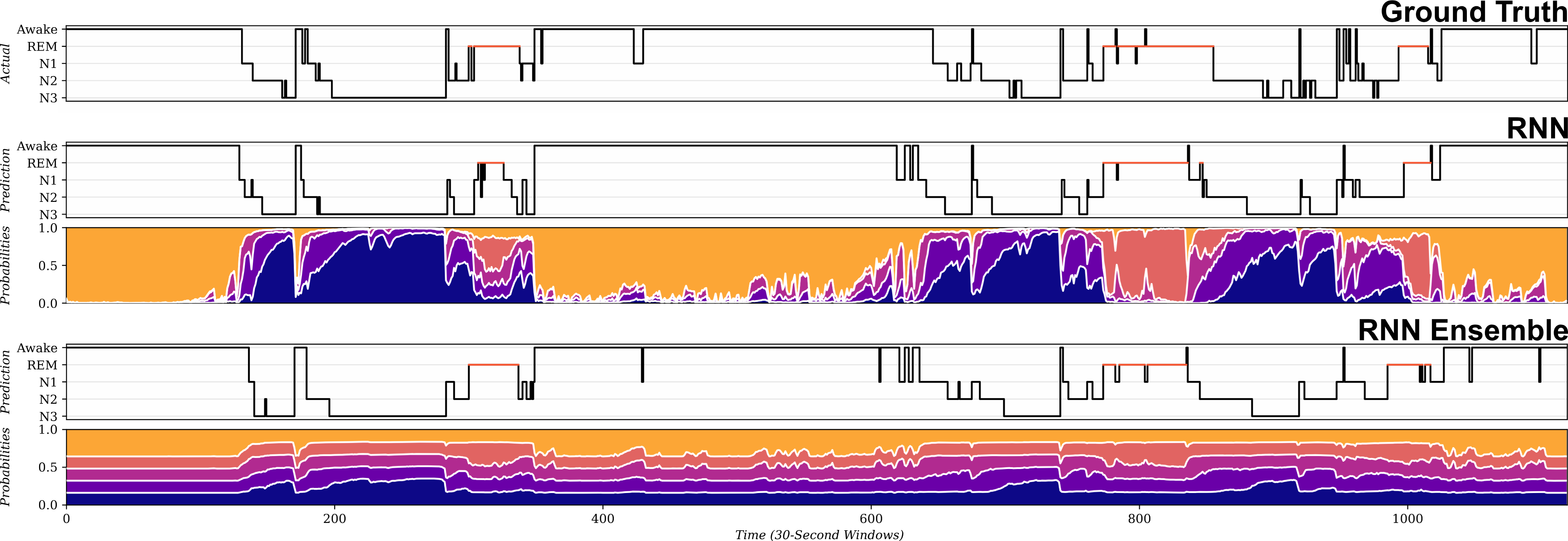}}
  \caption{Predicted sleep stages (hypnograms) and class probabilities (hypnodensities) for a typical single full-night recording in the test set for 5-class regular and ensemble models. The hypnogram prediction at each epoch is the class with the highest probability. Probabilities are coloured by ground truth sleep stage: darker to lighter colours represent classes as ordered in the hypnograms from bottom to top. The methods of calculating both hypnodensities can be found in \ref{sec:detailed-architecture}. An alternative method of calculating hypnodensities is shown in \ref{sec:ensemble-mode} (Figure \ref{fig:alternate-hypnodensity}).}
  \label{fig:monolith}
\end{figure}

Typical-case recording predictions for all 5-class models are shown in Figure \ref{fig:monolith} as hypnograms along with predicted point-wise output class probabilities, known as hypnodensity graphs \cite{anderer}. A Uniform Manifold Approximation and Projection (UMAP) \cite{umap} of an intermediate model output is also provided in \ref{sec:ensemble-mode} (Figure \ref{fig:umap}), which is able to show the separation of sleep stages as clusters in the learned high-dimensional representation space.

Figure \ref{fig:performance} displays the confusion matrices for all $n$-class ensemble models and overall metrics across all $n$-class regular and ensemble models.  We plotted performance in both 1-second and 30-second epochs and found similar results irrespective of the epoch length (balanced accuracies of 1-second RNN ensembles are 83.74\%, 75.06\%, 64.86\% for 3/4/5-class models).  Accordingly, all results in this paper are for 30-second windows.  Compared to a base RNN model, the ensembling procedure resulted in improved overall performance (5-class balanced accuracies of 86.19\%, 69.62\%, 67.72\%, 81.16\%, and 86.60\% for wake, N1, N2, N3, and REM sleep respectively, \ref{sec:classwise-metrics}). Most misclassifications were made between N1 and every other class, as well as between N2 and N3 for 5-class sleep staging. 
\pagebreak


\begin{figure}[H]
\centerline{\includegraphics[width=1.15\textwidth]{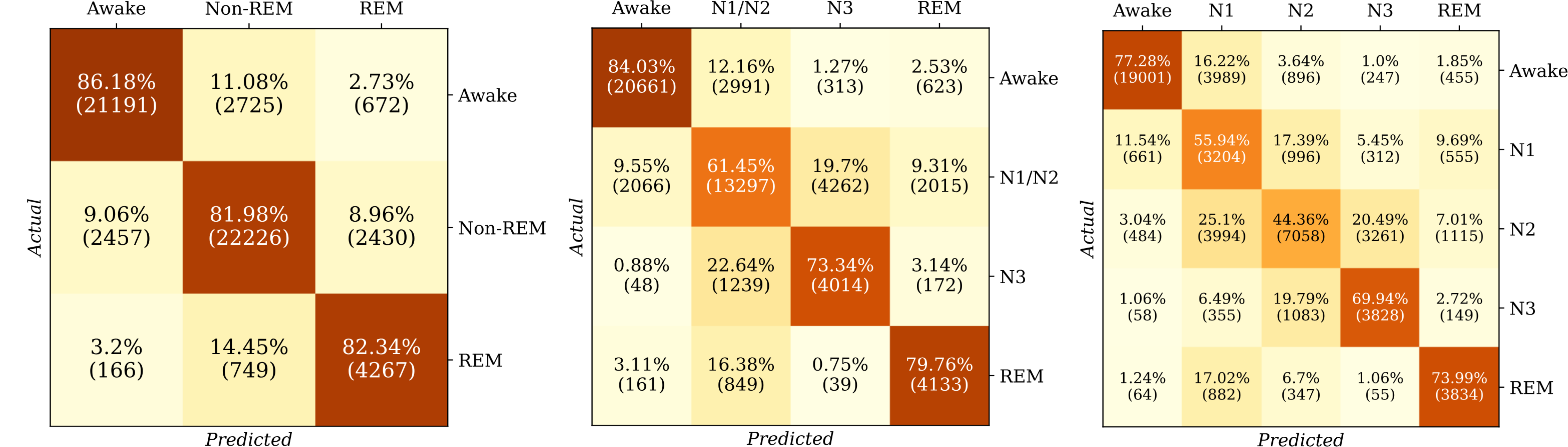}}
\end{figure}

\vspace{-6mm}

\begin{table}[H]
\footnotesize
\centerline{\begin{tabular}{cr|cccccc}
\Xhline{3\arrayrulewidth} 
\rowcolor[HTML]{EFEFEF} 
\cellcolor[HTML]{EFEFEF} & \multicolumn{1}{c|}{\cellcolor[HTML]{EFEFEF}} & \multicolumn{6}{c}{\cellcolor[HTML]{EFEFEF}\textit{\phantom{\Large{A}}Metric (\%)\phantom{\Large{A}}}} \\
\rowcolor[HTML]{EFEFEF} 
\multirow{-2}{*}{\cellcolor[HTML]{EFEFEF}\textit{Classes}} & \multicolumn{1}{r|}{\multirow{-2}{*}{\cellcolor[HTML]{EFEFEF}\textit{Model}}} & \multicolumn{1}{c}{\cellcolor[HTML]{EFEFEF}Balanced Accuracy} & \multicolumn{1}{c}{\cellcolor[HTML]{EFEFEF}Precision} & \multicolumn{1}{c}{\cellcolor[HTML]{EFEFEF}Recall} & \multicolumn{1}{c}{\cellcolor[HTML]{EFEFEF}F1} & \multicolumn{1}{c}{\cellcolor[HTML]{EFEFEF}Cohen's $\kappa$} & \multicolumn{1}{c}{\cellcolor[HTML]{EFEFEF}MCC} \\ \hline

 & \phantom{\Large{A}} RNN & 60.20 & 69.48 & 61.20 & 63.68 & 49.16 & 50.18 \\
 \multirow{-2}{*}{5} & RNN Ensemble & \textbf{65.11} & \textbf{71.61} & \textbf{64.41} & \textbf{66.15} & \textbf{53.23} & \textbf{54.38} \\ \hdashline[1pt/1pt]

 & \phantom{\Large{A}} RNN & 67.59 & 73.08 & 66.59 & 68.34 & 51.95 & 53.06 \\
\multirow{-2}{*}{4} & RNN Ensemble & \textbf{75.30} & \textbf{75.80} & \textbf{73.68} & \textbf{74.10} & \textbf{61.51} & \textbf{61.95} \\ \hdashline[1pt/1pt]

 & \phantom{\Large{A}} RNN & 81.09 & 83.20 & 81.98 & 82.35 & 69.59 & 69.77 \\
\multirow{-2}{*}{3} & RNN Ensemble & \textbf{84.02} & \textbf{84.76} & \textbf{84.04} & \textbf{84.23} & \textbf{72.89} & \textbf{73.00} \\ 

\Xhline{3\arrayrulewidth} 
\end{tabular}
}
\end{table}

\vspace{-6mm} 

\begin{figure}[H]
  \caption{Test set confusion matrices of ensembled RNN model for 3, 4, and 5 class sleep staging (top); Macro-evaluation metrics for all $n$-class models on the test set (bottom), where the best-performing model for each metric is bolded.}
  \label{fig:performance}
\end{figure}

In a set of young and healthy sleepers (age < 40, AHI < 5, PLMI < 5), the RNN ensemble model performed even better as denoted in Figure \ref{fig:healthy-subset} although removal of these participants from the test set did not substantially worsen model performance (\ref{sec:additional-metrics}, Figure \ref{fig:unhealthy-subset}), indicating that overall model performance was not driven by this small number of healthy participants. 

\begin{figure}[H]
\centerline{\includegraphics[width=1.15\textwidth]{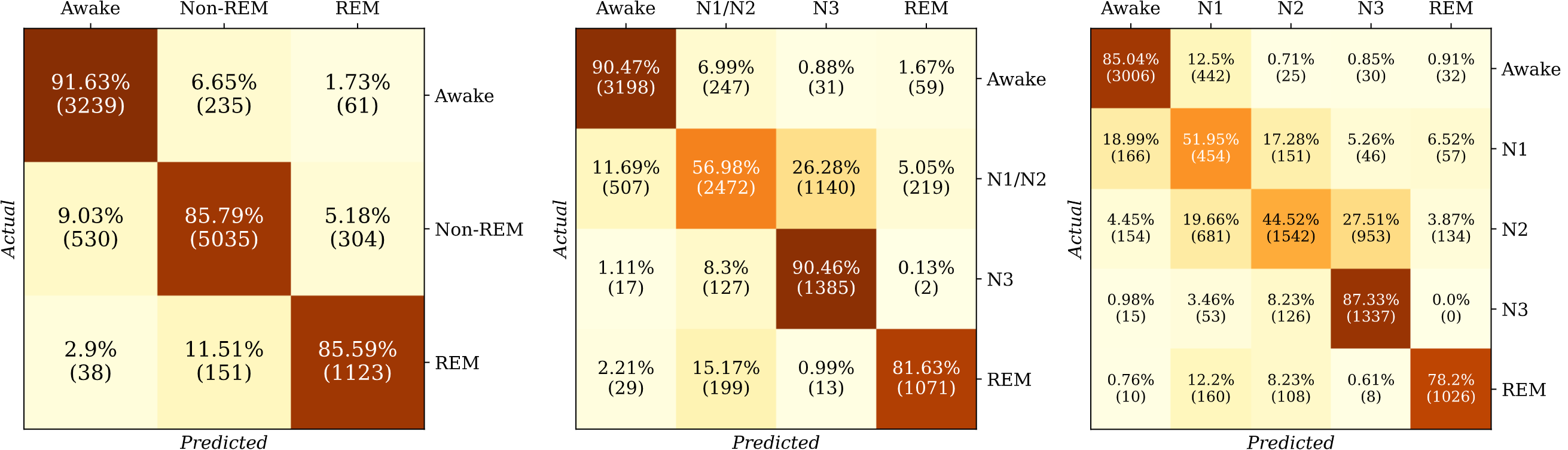}}
\end{figure}

\vspace{-6mm}

\begin{table}[H]
\footnotesize
\centerline{\begin{tabular}{c|cccccc}
\Xhline{3\arrayrulewidth} 
\rowcolor[HTML]{EFEFEF} 
\cellcolor[HTML]{EFEFEF} & \multicolumn{6}{c}{\cellcolor[HTML]{EFEFEF}\phantom{\Large{A}}\textit{Metric (\%)}\phantom{\Large{A}}} \\

\rowcolor[HTML]{EFEFEF} 
\multirow{-2}{*}{\cellcolor[HTML]{EFEFEF}\textit{Classes}} & Balanced Accuracy & Precision & Recall & F1 & Cohen's $\kappa$ & MCC \\ \Xhline{3\arrayrulewidth} 

\phantom{\Large{A}}5\phantom{\Large{A}} & 69.41 & 75.30 & 68.73 & 69.58 & 59.83 & 61.17 \\

4 & 79.89 & 78.40 & 75.83 & 75.57 & 66.42 & 67.58 \\

3 & 87.67 & 88.18 & 87.69 & 87.78 & 79.09 & 79.29 \\ \Xhline{3\arrayrulewidth} 

\end{tabular}}
\end{table}

\vspace{-6mm}

\begin{figure}[H]
\caption{Confusion matrices and macro-evaluation metrics of 3-class, 4-class, and 5-class RNN ensemble models on a healthy subset of the test set ($n$ = 11) defined as age < 40 and AHI < 5 and PLMI < 5.}
  \label{fig:healthy-subset}
\end{figure}

\vspace{-7mm}




\subsection{Feature Ablation Study}

We next investigated which sets of sensors (chest module vs.\ limb module; ECG vs.\ PPG vs.\ other) were most influential to model performance. Modified instances of the ensemble model were trained to determine the utility of its components. Firstly, features solely from the ANNE One’s chest module (ECG, accelerometry, chest temperature) or limb module (PPG, limb temperature) were examined. As shown in Table~\ref{tab:ablation}, the chest-only model trails closely behind the original while the limb-only model exhibits a marked degradation ($\approx10\%$) for all metrics. This strongly suggests that the chest module is the principal driver behind model accuracy, although the limb module alone is capable of recovering sleep staging information to a rough degree. Next, because the ECG and PPG sensors themselves are the most likely to experience poor signal quality, we examined the performance of a model trained on only the accelerometry and temperature sensors. Remarkably, exclusion of ECG and PPG data resulted in minimal loss of accuracy, meaning that information in the ECG and PPG is also being captured effectively by the accelerometry signals, with temperature providing additional complementary information. Further exclusion of temperature data resulted in an accelerometry-only model that performed similarly to or even slightly better (Table~\ref{tab:accl-ablation}), indicating that chest accelerometry alone captures most of the discriminative information used by the network. This is corroborated by a per-sensor feature importance analysis with Integrated Gradients \cite{sundararajan2017axiomatic} (Figure \ref{fig:feature-importance}), which suggests that accelerometry-derived features contribute the most to model performance relative to temperature, PPG, and ECG-derived features.


\vspace{-3mm}

\begin{table}[H]
\caption{Evaluation metrics for the 3-class, 4-class, and 5-class RNN ensemble models using different subsets of the entire feature set. The best-performing model for each metric is bolded.}
\label{tab:ablation}
\footnotesize
\centerline{\begin{tabular}{cr|cccccc}
\Xhline{3\arrayrulewidth} 
\rowcolor[HTML]{EFEFEF} 
\cellcolor[HTML]{EFEFEF} & \cellcolor[HTML]{EFEFEF} & \multicolumn{6}{c}{\cellcolor[HTML]{EFEFEF}\phantom{\Large{A}}\textit{Metric (\%)}\phantom{\Large{A}}} \\
\rowcolor[HTML]{EFEFEF} 
\multirow{-2}{*}{\cellcolor[HTML]{EFEFEF}\textit{Classes}} & \multirow{-2}{*}{\cellcolor[HTML]{EFEFEF}\textit{Feature Set}} & Balanced Accuracy & Precision & Recall & F1 & Cohen's $\kappa$ & MCC \\ \Xhline{3\arrayrulewidth} 

 & \phantom{\Large{A}}Full & \textbf{65.11} & \textbf{71.61} & \textbf{64.41} & \textbf{66.15} & \textbf{53.23} & \textbf{54.38} \\
 & Chest Module & 63.94 & 70.66 & 62.63 & 64.22 & 51.23 & 52.59 \\
 & Limb Module & 54.83 & 62.29 & 55.08 & 56.56 & 41.60 & 42.59 \\
\multirow{-4}{*}{5} & No ECG/PPG & 63.85 & 70.68 & 63.35 & 65.03 & 51.97 & 53.12 \\
\hdashline[1pt/1pt]

 & \phantom{\Large{A}}Full & \textbf{75.30} & \textbf{75.80} & \textbf{73.68} & \textbf{74.10} & \textbf{61.51} & \textbf{61.95} \\
 & Chest Module & 73.46 & 74.63 & 71.64 & 72.26 & 58.88 & 59.47 \\
 & Limb Module & 64.89 & 68.63 & 63.59 & 64.66 & 48.22 & 49.13 \\
\multirow{-4}{*}{4} & No ECG/PPG & 73.03 & 74.46 & 71.52 & 72.10 & 58.70 & 59.30 \\
\hdashline[1pt/1pt]

 & \phantom{\Large{A}}Full & \textbf{84.02} & \textbf{84.76} & \textbf{84.04} & \textbf{84.23} & \textbf{72.89} & \textbf{73.00} \\
 & Chest Module & 81.77 & 83.58 & 82.50 & 82.83 & 70.39 & 70.53 \\
 & Limb Module & 73.28 & 78.10 & 74.71 & 75.84 & 58.21 & 58.69 \\
\multirow{-4}{*}{3} & No ECG/PPG & 82.78 & 84.26 & 83.18 & 83.51 & 71.71 & 71.56 \\
\Xhline{3\arrayrulewidth} 

\end{tabular}}
\end{table}

\vspace{-3mm}

To better characterize the information available in chest accelerometry, we probed the signals within these channels. At the individual-participant level, aggregated accelerometry spectrograms aligned to the hypnogram (Section~\ref{sec:hypno-spectrogram}) show stage-dependent structure, with clear shifts in power concentrated in a low-frequency band overlapping with the range of expected respiratory rates in sleep. This supports the hypothesis that very low-frequency components of chest accelerometry are important for distinguishing sleep stages. To examine this quantitatively, we (1) trained a model using accelerometry channels alone and then (2) replaced the band-pass filter with a high-pass cutoff to $0.5$ Hz to attenuate the expected frequency range of resting respirations. Results are shown in Table \ref{tab:accl-ablation}. The accelerometry-only model performed marginally better than the No ECG/PPG model but worse than the full model; this may be attributable to the data's unimodal nature, which could reduce the complexity of learning structural patterns in multiple data types during training, thus improving convergence. Meanwhile, the reduced performance of the high-passed accelerometry-only model is compatible with the possibility that some information used by the model is in the frequency range of human respirations.


\subsection{Alternative Approaches}


To guard against overfitting and reduce the computation burden, we downsampled the ECG, PPG, and accelerometry signals to 100Hz prior to use in our model. To explore the extent to which this would have impacted performance, we retrained the model without downsamping (i.e. using 512 Hz ECG, 128 Hz PPG, and 210 Hz accelerometry data) and band-pass filtering. This did not result in any decisive improvement in model performance (\ref{sec:additional-metrics}, Figure \ref{fig:extra-conf-matrices}). A strong correlation between high-frequency and low-frequency bins might have contributed to redundancy and increased the likelihood of overfitting to noise. In support of this hypothesis, we note the difference in balanced accuracy between the training and test sets was much less in the downsampled data ($\approx 5\%$) than in the non-downsampled data ($\approx 10\%$).  

Two variants of the RNN model were produced to determine the value of the Mamba architecture and the 1-second window approach. The first, a LSTM-based RNN, replaces the Mamba blocks with LSTM blocks of identical input, hidden, and output dimension sizes, resulting in a comparable parameter count to the original RNN model. Following our previous study \cite{sleep_poster}, the second is a CRNN ($\approx$1271K parameters) that groups every 30 time windows into buckets, extracts bucket features with a CNN, and processes features sequentially using a Mamba-based RNN, resulting in predictions made per 30 seconds (see \ref{sec:crnn-info} for details). The technique of extracting features from 30 seconds of data for an autoregressive model has seen many variants \cite{sun, sleepppg, zhang2018sleep}, making the CRNN suitable as a "baseline" approach to windowing. As provided by Table \ref{tab:additional_architectures} (confusion matrices in \ref{sec:additional-metrics}, Figure \ref{fig:extra-conf-matrices}), both the LSTM-based RNN models and the CRNN models have modestly reduced performance in all metrics, confirming the Mamba approach and 1-second window approach are beneficial.

\vspace{-2mm}







\begin{table}[H]
\caption{Evaluation metics for the 3-class, 4-class, and 5-class ensemble RNN, LSTM, and CRNN models. The best-performing model for each metric is bolded.}
\label{tab:additional_architectures}
\footnotesize
\centerline{\begin{tabular}{cr|cccccc}
\Xhline{3\arrayrulewidth} 
\rowcolor[HTML]{EFEFEF} 
\cellcolor[HTML]{EFEFEF} & \multicolumn{1}{c|}{\cellcolor[HTML]{EFEFEF}} & \multicolumn{6}{c}{\cellcolor[HTML]{EFEFEF}\textit{\phantom{\Large{A}}Metric (\%)\phantom{\Large{A}}}} \\
\rowcolor[HTML]{EFEFEF} 
\multirow{-2}{*}{\cellcolor[HTML]{EFEFEF}\textit{Classes}} & \multicolumn{1}{r|}{\multirow{-2}{*}{\cellcolor[HTML]{EFEFEF}\textit{Model}}} & \multicolumn{1}{c}{\cellcolor[HTML]{EFEFEF}Balanced Accuracy} & \multicolumn{1}{c}{\cellcolor[HTML]{EFEFEF}Precision} & \multicolumn{1}{c}{\cellcolor[HTML]{EFEFEF}Recall} & \multicolumn{1}{c}{\cellcolor[HTML]{EFEFEF}F1} & \multicolumn{1}{c}{\cellcolor[HTML]{EFEFEF}Cohen's $\kappa$} & \multicolumn{1}{c}{\cellcolor[HTML]{EFEFEF}MCC} \\ \hline

 
 & \phantom{\Large{A}}CRNN & 60.02 & 56.30 & 60.02 & 56.37 & 46.96 & 47.92 \\
5 & LSTM-Based RNN & 62.64 & 69.46 & 62.47 & 64.03 & 50.76 & 51.87 \\ 
 & RNN & \textbf{65.11} & \textbf{71.61} & \textbf{64.41} & \textbf{66.15} & \textbf{53.23} & \textbf{54.38} \\ \hdashline[1pt/1pt]

 & \phantom{\Large{A}}CRNN & 70.59 & 63.68 & 70.59 & 66.09 & 55.13 & 55.47 \\
4 & LSTM-Based RNN & 72.82 & 73.95 & 71.64 & 72.06 & 58.70 & 59.17 \\ 
 & RNN & \textbf{75.30} & \textbf{75.80} & \textbf{73.68} & \textbf{74.10} & \textbf{61.51} & \textbf{61.95} \\ \hdashline[1pt/1pt]

 & \phantom{\Large{A}}CRNN & 80.21 & 73.43 & 80.21 & 75.76 & 65.26 & 65.46 \\
3 & LSTM-Based RNN & 81.06 & 82.90 & 81.37 & 81.82 & 68.71 & 68.95 \\ 
 & RNN & \textbf{84.02} & \textbf{84.76} & \textbf{84.04} & \textbf{84.23} & \textbf{72.89} & \textbf{73.00} \\ 

\Xhline{3\arrayrulewidth} 
\end{tabular}
}
\end{table}

\subsection{Global Sleep Metrics}

\begin{figure}[H]
\centerline{\includegraphics[width=1.15\textwidth]{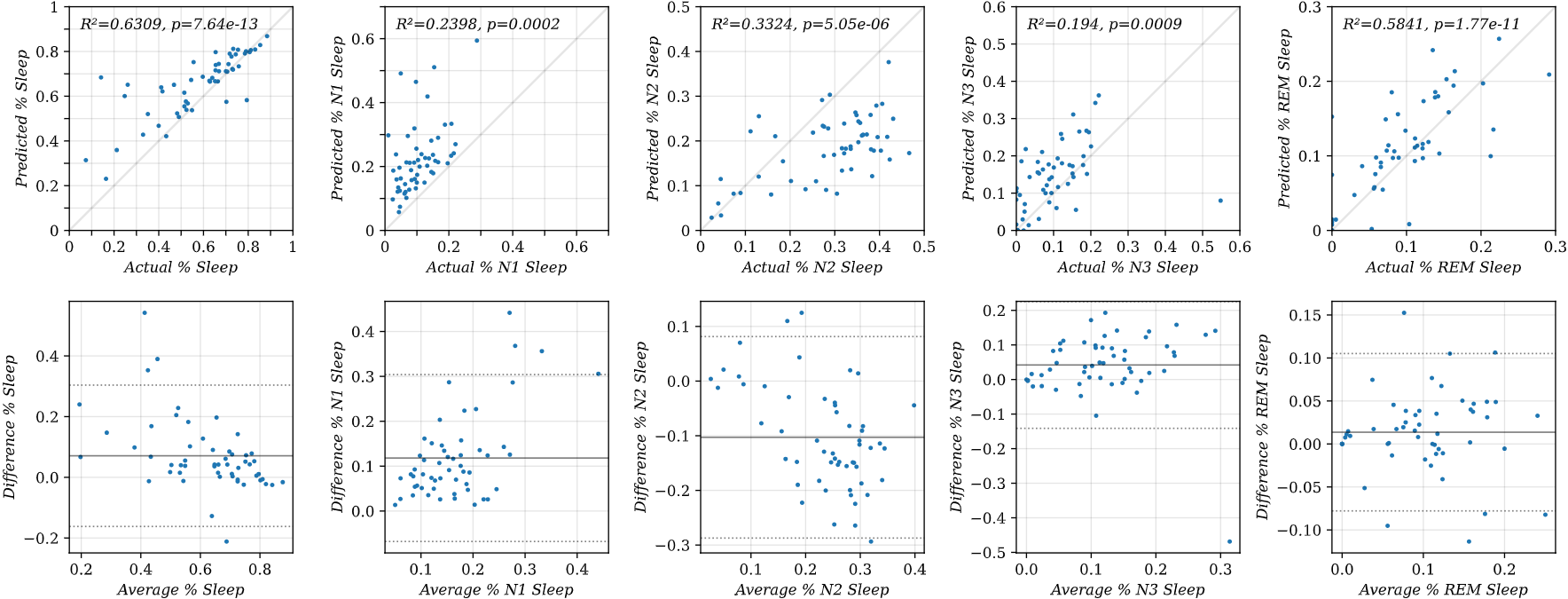}}
  \caption{Predicted vs. actual sleep stage scatter plots (top row) and Bland-Altman plots (bottom row) for the 5-class RNN ensemble model on the test set. The center horizontal line is the mean. The two dotted lines are 1.96 standard deviations above and below the mean.}
  \label{fig:bland-altman}
\end{figure}

As a means of evaluating model bias and verifying model robustness, Figure \ref{fig:bland-altman} contains scatter plots and Bland-Altman plots for the 5-class RNN ensemble, comparing predicted vs. actual values for a number of summary metrics including time asleep, \% N1 sleep, \% N2 sleep, \% N3 sleep, and \% REM sleep.  The model trends towards slightly underestimating N2 sleep and wake (as shown by the overestimated sleep in the first column) and slightly overestimating N1 and N3 sleep.

\pagebreak
\subsection{Recording-Wise Performance}
We computed recording-wise performance metrics for the 5-class RNN ensemble model, the distributions of which are summarized in Table \ref{tab:results-by-group} . We then considered recording-level accuracy as a function of clinical and demographic features, and of ECG and PPG signal quality, using linear regression models or $t$-tests (plots in \ref{sec:additional-metrics}, Figures \ref{fig:continuous-plots}, \ref{fig:sqi-plot}, \ref{fig:disorders}). The model performed similarly on recordings from individuals with insomnia as compared to those without (accuracy $+0.03$ for those with as compared to without insomnia, $p=0.42$) and in those with RLS compared to those without (accuracy $+0.03$, $p=0.40$). 

Accuracy sunk for those with REM sleep behaviour disorder (accuracy $-0.12$, $p=0.02$), largely driven by poorer classification of REM sleep (39\% and 79\% of REM sleep was detected by the 5-class RNN ensemble for those with and without RBD respectively). Accuracy also declined with higher AHI ($p=0.04$); however, this was a modest effect, with accuracy dropping only 2\% for every 10 unit higher AHI. Accuracy was not significantly associated with age, sex, periodic limb movement index, sleep efficiency, or total sleep time.  

Model accuracy did not vary significantly with the proportion of the recording with bad quality ECG (estimate $-0.1$\% decline in accuracy for each 1\% increase in the proportion of poor-quality ECG, $p=0.07$) or PPG (estimate $-0.01$\% decline in accuracy per 1\% increase in the proportion of poor-quality PPG, $p=0.88$), but it was significantly worse in recordings with a high proportion of epochs with neither good ECG nor good PPG SQI (estimate $-0.3$\% decline in accuracy for each 1\% increase in the proportion of time where both ECG and PPG SQI were poor, $p=0.03$). However, this effect was modest ($\text{R}^2=0.09$) and we note that the proportion of time where this occurred was generally low, with all but 6 recordings having at least one of ECG or PPG with good quality for at least 90\% of the recording. Even in recordings in which 40\% or more of the recording had neither good quality ECG nor good quality PPG, 5-class model accuracy remained above 0.50.

Four annotators were involved in generating the ground truth labels used in this study and the agreement between the model and the ground truth labels did not differ by the individual who generated the annotations ($p = 0.89$, one-way F-test).

\vspace{-3mm}

\begin{table}[H]
\caption{Per-recording accuracy for the 5-class RNN ensemble model (top); $\text{R}^2$ and slope/$t$-test $p$-values of a line fitting various sleep (middle) and SQI statistics (bottom) to accuracy using linear regression. Statistically significant $p$-values (< 0.05) are bolded.}
\label{tab:results-by-group}
\centering
\small
\begin{tabular}{l|rr}
\Xhline{3\arrayrulewidth}

\rowcolor[HTML]{EFEFEF}
\multicolumn{1}{c|}{\cellcolor[HTML]{EFEFEF}{\phantom{\Large{A}}\textit{Metric (\%)}\phantom{\Large{A}}}} & \multicolumn{1}{c}{\cellcolor[HTML]{EFEFEF}$\mu$} & \multicolumn{1}{c}{\cellcolor[HTML]{EFEFEF}$\sigma$} \\ \hline

Accuracy\phantom{\Large{A}} & 64.54 & 10.77 \\
F1 & 66.72 & 10.28 \\
Cohen's $\kappa$ & 50.58 & 14.50 \\
Matthews correlation coefficient & 52.92 & 13.40 \\ \Xhline{3\arrayrulewidth}

\rowcolor[HTML]{EFEFEF} 
\multicolumn{1}{c|}{\textit{Sleep Statistic}} & \multicolumn{1}{c}{$\text{R}^2$} & \multicolumn{1}{c}{\phantom{\Large{A}}$p$\phantom{\Large{A}}} \\ \hline

Age\phantom{\Large{A}} & 0.0618 & 0.0700 \\
Sex & 0.0334 & 0.1861 \\
Periodic Limb Movement Index & 0.0338 & 0.1830 \\
Apnea-Hypopnea Index & 0.0751 & \textbf{0.0449} \\
Body Mass Index & 0.0350 & 0.1753 \\
Total Sleep Time & 0.0055 & 0.5940 \\
Sleep Efficiency & 0.0002 & 0.9281 \\ \Xhline{3\arrayrulewidth}

\rowcolor[HTML]{EFEFEF} 
\multicolumn{1}{c|}{\textit{Disorder $(n)$}} & \multicolumn{1}{c}{$\text{R}^2$} & \multicolumn{1}{c}{\phantom{\Large{A}}$p$\phantom{\Large{A}}} \\ \hline

Restless Leg Syndrome (9)\phantom{\Large{A}} & 0.0134 & 0.4044 \\
Insomnia (14) & 0.0135 & 0.4032 \\
REM Sleep Behaviour Disorder (5) & 0.0965 & \textbf{0.0222} \\ \Xhline{3\arrayrulewidth}

\rowcolor[HTML]{EFEFEF} 
\multicolumn{1}{c|}{\textit{\% of Recording with Good SQI In...}} & \multicolumn{1}{c}{$\text{R}^2$} & \multicolumn{1}{c}{\phantom{\Large{A}}$p$\phantom{\Large{A}}} \\ \hline

ECG\phantom{\Large{A}} & 0.0623 & 0.0687 \\
PPG & 0.0005 & 0.8753 \\
ECG or PPG & 0.0913 & \textbf{0.0264} \\
ECG and PPG & 0.0136 & 0.4017 \\ \Xhline{3\arrayrulewidth}
\end{tabular}
\end{table}

\section{Discussion}

\subsection{Key Findings}




This work has developed a Mamba-based RNN ensemble model to accurately perform inference from the ANNE One device for major sleep stages. The model demonstrates good performance across a wide range of patient characteristics, without a need to screen and exclude recordings and epochs on the basis of signal quality.

\subsection{Related Studies}

Previous studies have undertaken the challenge of inferring sleep stage from non-EEG signals (Tables \ref{tab:related-works-PSG} \& \ref{tab:related-works-wearable}) using a mixture of cardiopulmonary (ECG, PPG, respiratory impedance bands, respiratory flow cannulas) and accelerometry data.  Many studies used data from conventional PSG recordings \cite{radha2019sleep, sun, sleepppg, pini, topalidis2023pulses, jones2024expert, kazemi2024improved} although a handful of studies used data from purpose-designed wearable sensors \cite{zhang2018sleep, sun, anne_personal, pini, topalidis2023pulses, silva, wulterkens:2021}.  In general, models trained on and applied to PSG data performed better than models utilizing wearable sensor data.  This is not surprising since signal quality is often much higher for attended in-laboratory PSGs where more intrusive sensors can be used,  signal quality is continuously monitored and issues can be addressed by the technologist as they occur.  Even so, despite using wearable rather than PSG sensor data, the macro performance of our ensemble RNN model approached \cite{sun, topalidis2023pulses, jones2024expert, radha2019sleep, sleepppg} or exceeded \cite{pini, kazemi2024improved} the performance reported for models trained on much larger sets of PSG data.   

Our model exceeded \cite{silva, pini, anne_personal, sun, zhang2018sleep} or matched \cite{wulterkens:2021, schipper} the benchmarks of most wearable sensor based models with the exception of one \cite{topalidis2023pulses}.  Of note, this study examined a younger set of participants (mean age 45) who were free of psychiatric and neurological co-morbidity, which may influence the relationship between sleep stage and its autonomic, cardiac, and pulmonary manifestations. In contrast, we studied an older (mean age 57.49) set of participants with significant sleep co-morbidity (45.81\% with AHI > 5).  When we limit our analyses to younger individuals without sleep apnea or significant periodic limb movements, model performance improves further (Figure \ref{fig:healthy-subset}). Another contrast between \cite{topalidis2023pulses} and the present study is that \cite{topalidis2023pulses} performed much stricter QC for recording quality with 54/136 or nearly 40\% of nights of wearable sensor recording excluded from analysis due to poor signal quality.  In contrast, in the interest of evaluating real-world performance, we did not exclude any recording or any epochs on the basis of signal quality and the reported results reflect consideration of all available epochs of recording, irrespective of signal quality. The robustness of our model to missing data or poor signal quality is likely related in part to its use of multiple data streams, in particular accelerometry, which appeared to capture much of the information that would otherwise have been captured by ECG or PPG, such that sensors returning poor-quality data are compensated for in the model by those returning good quality data. Indeed, poor signal quality was present in at least one sensor module for 866 (29.7\%) of the 2914 hours of recording in our dataset.  However, for 725 (83.7\%) of these hours, the other module returned sufficient quality signal to allow accurate predictions.  

Of particular comparative interest is the one previous paper addressing sleep stage inference from ANNE One data \cite{anne_personal}.  Ultimately, success in developing a population model was limited, and while there was some success in generating person-specific models, the performance of our ensemble RNN model exceeded even the best person-specific model in the paper which had a 4-class balanced accuracy of 73.4\%.  




\begin{table}[H]
\caption{Sleep staging models in related literature trained and evaluated on PSG data. For papers with confusion matrices, additional metrics were manually calculated and included in tandem with those explicitly reported by the paper. Acc. = accuracy, B.Acc. = balanced accuracy, $\kappa$ = Cohen's kappa.}
\label{tab:related-works-PSG}
\renewcommand{\arraystretch}{1.5}
\footnotesize
\centerline{\begin{tabular}{lllll}
    \rowcolor[HTML]{EFEFEF}
    \Xhline{3\arrayrulewidth}
    
    \textit{Study (Year)}\phantom{\Large{A}} & \textit{Dataset} & \textit{Metric} &  \textit{\%} &\textit{Signal Features} \\
    
    \Xhline{3\arrayrulewidth}
    
    \makecell[l]{{Radha et al. (2019) \cite{radha2019sleep}}} & \makecell[l]{SIESTA (7 sleep labratories)\\ $-$ 584 recordings (292 subjects) \\ $-$ 25\% testing} & \makecell[l]{4-stage Acc.\\ 4-stage $\kappa$} & \makecell[l]{77 \\ 61} & ECG \\ \hdashline[1pt/1pt]

    \makecell[l]{{Sun et al. (2020) \cite{sun}}} & \makecell[l]{MGH (sleep laboratory) \\ $-$ 8682 recordings (7208 subjects) \\ $-$ 20\% testing} & \makecell[l]{3-stage B.Acc.\\ 3-stage F1\\ 3-stage $\kappa$\\ 5-stage B.Acc. \\ 5-stage F1 \\ 5-stage $\kappa$} & \makecell[l]{86.8 \\ 84.2 \\ 76.0 \\ 70.3 \\ 68.1 \\ 58.5} & \makecell[l]{ECG,\\Resp. Effort}\\ \hdashline[1pt/1pt]
    
    \makecell[l]{{Kotzen et al. (2022) \cite{sleepppg}}} & \makecell[l]{SHHS Visit 1 (multi-center) \\ $-$ 5758 recordings (pretraining) \\ MESA (general population) \\ $-$ 2056 recordings \\ $-$ 204 testing} & \makecell[l]{4-stage Acc. \\ 4-stage B.Acc.\\ 4-stage F1\\ 4-stage $\kappa$} & \makecell[l]{84 \\ 76 \\ 83 \\ 75}&PPG \\ \hdashline[1pt/1pt]

    \makecell[l]{{Kotzen et al. (2022) \cite{sleepppg}}} & \makecell[l]{SHHS Visit 1 (multi-center) \\ $-$ 5758 recordings (pretraining) \\ CFS Visit-5 v1 (family study) \\ $-$ 324 recordings \\ $-$ 80 testing} & \makecell[l]{4-stage Acc. \\ 4-stage B.Acc.\\ 4-stage F1\\ 4-stage $\kappa$} & \makecell[l]{82 \\ 80 \\ 82 \\ 74}&PPG \\ \hdashline[1pt/1pt]

    \makecell[l]{{Pini et al. (2022) \cite{pini}}} & \makecell[l]{Unspecified (academic sleep centers) \\ $-$ 12404 recordings \\ CinC (sleep laboratory) \\ $-$ 994 recordings \\ $-$ External testing set} & \makecell[l]{3-stage B.Acc.\\ 3-stage F1\\ 3-stage $\kappa$\\ 3-stage MCC \\ 4-stage B.Acc.\\ 4-stage F1 \\ 4-stage $\kappa$\\ 4-stage MCC} & \makecell[l]{74.0 \\ 81.4 \\ 61.6 \\ 61.7 \\ 63.6 \\ 72.0 \\ 53.6 \\ 54.0} &  ECG \\ \hdashline[1pt/1pt]

    \makecell[l]{{Topalidis et al. (2023) \cite{topalidis2023pulses}}} & \makecell[l]{Unspecified (general population \\ w/ sleep complaints)\\$-$ 314 recordings (185 subjects)\\$-$ 25\% testing \\ $-$ Ambulatory home recordings}  & \makecell[l]{4-stage B.Acc\\ 4-stage F1 \\ 4-stage $\kappa$\\ 4-stage MCC} & \makecell[l]{85.4 \\ 86.3 \\ 79.2 \\ 79.2} & ECG \\
    \hdashline[1pt/1pt]

    \makecell[l]{{Jones et al. (2024) \cite{jones2024expert}}} & \makecell[l]{CCSHS, CFS, CHAT, MESA, WSC \\ $-$ 4000 random sampled recordings \\ $-$ 500 testing} & \makecell[l]{5-stage B.Acc\\ 5-stage Median $\kappa$} & \makecell[l]{74 \\ 72.5} & ECG \\ \hdashline[1pt/1pt]

    \makecell[l]{{Kazemi et al. (2024) \cite{kazemi2024improved}}} & \makecell[l]{UC Irvine Sleep Center \\ $-$ 123 recordings \\ $-$ 20\% testing} & \makecell[l]{3-stage $\kappa$ \\ 4-stage $\kappa$ \\ 5-stage $\kappa$}& \makecell[l]{71.4 \\ 55.0 \\ 61.6} & \makecell[l]{PPG, \\Resp. Flow Rate,\\Resp. Effort }\\

    \Xhline{3\arrayrulewidth}
  \end{tabular}}
\end{table}

\begin{table}[H]
\caption{Sleep staging models in related literature that are trained (unless otherwise specified) and evaluated on wearable data. For papers with confusion matrices, additional metrics were manually calculated and included in tandem with those explicitly reported by the paper. Resp. = respiratory, Accel. = accelerometry, Temp. = temperature, Acc. = accuracy, B.Acc. = balanced accuracy, Prec. = precision, Rec. = recall, $\kappa$ = Cohen's kappa, MCC = Matthews correlation coefficient.}
\label{tab:related-works-wearable}
\renewcommand{\arraystretch}{1.5}
\footnotesize
\centerline{\begin{tabular}{lllll}
    \rowcolor[HTML]{EFEFEF}
    \Xhline{3\arrayrulewidth}
    
    \textit{Study (Year)}\phantom{\Large{A}} & \textit{Dataset} & \textit{Metric} & \textit{\%} & \textit{Signal Features} \\

    \Xhline{3\arrayrulewidth}

   \makecell[l]{{Zhang et al. (2018) \cite{zhang2018sleep}}} & \makecell[l]{Beijing General Hospital of the Air Force \\ (general population) \\ $-$ 39 recordings \\ $-$ 12.5\% testing} & \makecell[l]{5-stage Prec.\\ 5-stage Rec.\\ 5-stage F1} & \makecell[l]{58.5 \\ 61.1 \\ 58.5} & \makecell[l]{Heart Rate,\\ Actigraphy} \\ \hdashline[1pt/1pt]

    \makecell[l]{{Sun et al. (2020) \cite{sun}} \\ \textit{Model from Table \ref{tab:related-works-PSG}}} & \makecell[l]{SHHS Visit 1 \& 2 (multi-center)\\ $-$ 1000 random sampled recordings \\ $-$ External testing set} & \makecell[l]{3-stage B.Acc. \\ 3-stage F1  \\ 3-stage $\kappa$ \\ 5-stage B.Acc.\\ 5-stage F1\\ 5-stage $\kappa$}& \makecell[l]{80.8 \\ 80.2 \\ 69.7 \\ 63.9 \\ 58.6 \\ 53.3} & \makecell[l]{ECG,\\Resp. Effort}\\ \hdashline[1pt/1pt]

    \makecell[l]{{Wulterkens et al. (2021) \cite{wulterkens:2021}}} & \makecell[l]{SOMNIA, N2N, HHS, HealthBed \\ $-$ 835 recordings \\ $-$ 35\% testing} & \makecell[l]{4-stage B.Acc.\\ 4-stage F1.\\ 4-stage $\kappa$ \\ 4-stage MCC} & \makecell[l]{75.4 \\ 76.4 \\ 63.7 \\ 63.7} & \makecell[l]{PPG,\\Accel.} \\ \hdashline[1pt/1pt]

    \makecell[l]{{Chen et al. (2022) \cite{anne_personal}} \\ \textit{Individual-Based Model}}  & \makecell[l]{Shirley Ryan AbilityLab (inpatient unit) \\ $-$ 10 recordings \\ $-$ 10\% testing} & \makecell[l]{3-class Mean $\kappa$ \\ 4-class B.Acc.\\ 4-class Mean $\kappa$} & \makecell[l]{60.0 \\ 73.0 \\ 61.7} & \makecell[l]{ECG, PPG,\\Accel., Temp.} \\ \hdashline[1pt/1pt]

    \makecell[l]{{Chen et al. (2022) \cite{anne_personal}} \\ \textit{Population Model}} & \makecell[l]{Shirley Ryan AbilityLab (inpatient unit) \\ $-$ 10 recordings \\ $-$ 10\% testing} & \makecell[l]{3-class Mean $\kappa$\\ 4-class Mean $\kappa$}& \makecell[l]{17.1 \\ \>\>6.1} & \makecell[l]{ECG, PPG,\\Accel., Temp.} \\ \hdashline[1pt/1pt]

    \makecell[l]{{Pini et al. (2022) \cite{pini}} \\ \textit{Model from Table \ref{tab:related-works-PSG}}} & \makecell[l]{Z3Pulse (general population) \\ $-$ 156 recordings (52 subjects) \\ $-$ ECG-based wearable \\ $-$ External testing set} & \makecell[l]{3-stage B.Acc.\\ 3-stage F1\\ 3-stage $\kappa$\\ 3-stage MCC\\ 4-stage B.Acc.\\ 4-stage F1\\ 4-stage $\kappa$\\ 4-stage MCC}& \makecell[l]{72.5 \\ 79.8 \\ 61.2 \\ 61.4 \\ 64.2 \\ 69.8 \\ 53.2 \\ 53.6} & ECG \\ \hdashline[1pt/1pt]
    
    \makecell[l]{{Topalidis et al. (2023) \cite{topalidis2023pulses}} \\ \textit{Model from Table \ref{tab:related-works-PSG}}} & \makecell[l]{Unspecified (general population w/ sleep \\ complaints)\\$-$ 314 recordings (185 subjects)\\$-$ 25\% testing}  & \makecell[l]{4-stage B.Acc \\ 4-stage F1 \\ 4-stage $\kappa$ \\ 4-stage MCC}& \makecell[l]{84.4 \\ 83.9 \\ 76.2 \\ 76.3} & ECG, PPG \\ \hdashline[1pt/1pt]

    \makecell[l]{Schipper et al. (2024) \cite{schipper}} & \makecell[l]{Sleep Medicine Center Kempenhaeghe,\\OLVG Hospital (at-home recordings)\\$-$ 323 recordings} & \makecell[l]{4-stage B.Acc.\\4-stage F1\\4-stage $\kappa$\\4-stage MCC} & \makecell[l]{75.6\\79.4\\67.1\\67.2} & \makecell[l]{Accel.}\\\hdashline[1pt/1pt]

    \makecell[l]{{Silva et al. (2024) \cite{silva}}} & \makecell[l]{Instituto do Sono (general population)\\ $-$ 1522 recordings (1430 subjects)\\$-$ 586 testing} & \makecell[l]{4-stage Mean Acc.\\ 4-stage Mean $\kappa$}& \makecell[l]{70.7 \\ 56} & \makecell[l]{Heart Rate,\\Accel.} \\

    \Xhline{3\arrayrulewidth}
  \end{tabular}}
\end{table}  


\subsection{Model Performance}

The RNN model architecture obtained maximal per-class performances for wake and REM, and minimal per-class performances for N1 and N2 sleep. As a transitory state, periods of N1 sleep are short and instable, making its discrimination a difficult task. Physiological similarities between N2 and N3 sleep, such as cardiorespiratory signatures, similarly complicate the model's ability to differentiate them, as evidenced by N2--N3 confusion as the most frequent misprediction for 5-class models (as well as N1/N2--N3 confusion for 4-class models). Of note, these classes also represent those hardest to distinguish from each other in sleep staging by human annotators \cite{rosenberg:2013}. Conversely, wake and REM sleep have clearer differentiating characteristics, such as higher heart rate, which allows for easier classification.

No statistically significant variations in accuracy based on age, sex, PLMI, BMI, TST, sleep efficiency, insomnia, or RLS were exhibited by the 5-class ensemble model, providing an important advantage when applied in clinical contexts. The marginally significant ($p=0.0449$) association between AHI and accuracy may owe to the fact that individuals with sleep apnea experience sleep fragmentation, resulting in N1 sleep, one of the most difficult stages to separate; nonetheless, the association was modest ($\text{R}^2=0.0751$). The model showed poorer performance in individuals with RBD, driven by difficulties in classifying REM sleep. This may reflect in part the centrality of accelerometry to our model, which, in the context of abnormal movements in REM sleep, may be apt to misclassify REM sleep.  

The series of ablation analyses establish that much of the RNN model's performance was likely fueled by chest module features (i.e.\ ECG, temperature, and XYZ accelerometry) rather than limb module (i.e.\ PPG and temperature), meaning that omitting the limb module may be reasonable if participant burden is a concern and collection of oximetry data is not necessary. Exclusion of both ECG and PPG data with retention of only accelerometry and temperature data resulted in model performance almost indistinguishable from use of all the chest module sensors, suggesting that the inclusion of traditional heart rate sensors (PPG, ECG) is not necessary. This highlights the richness of accelerometry data obtained at the chest, which may capture respiratory data as well as measures of movement, and may also contribute to the robustness of our model given the relative reliability of the accelerometers compared to the ECG/PPG sensors. At the same time, the attenuation of low-frequency activity (\ref{sec:accl-ablation}) produced only modest degradations in performance, suggesting that the model is exploiting a nuanced combination of low-frequency respiratory motion, postural changes, and other movements, rather than any single narrow frequency band alone.

\subsection{Limitations, Strengths, and Future Directions}

The core strengths of this study are: 1) the ANNE One system itself, with multiple channels of data recorded from a pair of minimally intrusive sensors, is a strength insofar as it provides an element of redundancy should one sensor return low quality signal, 2) the inclusion of Mamba allowed the achievement of relatively high performance with a relatively parameter-scarce architecture, and 3) the study population was large and varied, including participants from across a spectrum of sleep disorders.

A few methodological limitations are also worth considering. 1) The black box nature of the RNN makes it difficult to ascertain the specific mechanisms used to infer sleep stage from the sensor data. 2) The recordings were overnight with a brief period of wakefulness before lights out, resulting in a preponderance of sleep vs.\ wake, so it is difficult to know for sure how the model will perform in detecting daytime naps during long ambulatory recordings, and further studies with ambulatory recordings in clinical populations will be informative. 3) The recordings were all obtained from a single site, and additional work is needed to assess the extent to which these results will be generalizable to ANNE recordings made in different settings on different patient populations. 

The choice of 1-second windows created both advantages and disadvantages. It reduced the number of frequency bins per window, saving on model parameter count, and inflated the number of epochs by thirtyfold, which Mamba's long-term dependencies could effectively capture. However, this setup complicated interpretability: while high-frequency content ($>1$ Hz) emerged in the model's frequency bin features, low-frequency content ($<1$ Hz) manifested as changes across time windows, making their effects more difficult to capture with feature importance techniques.


An important future direction will be to better delineate how chest accelerometry contributes so strongly to model robustness, particularly in the absence of ECG and PPG. Our results suggest that some of this may be related to low-frequency, possibly respiratory, activity.  However, the model appears to be making use of other information as well, and defining the physiological basis of this will be critical to better understanding the potential and limitations of chest accelerometry-based sleep staging.

\section{Conclusion}


This work introduces a Mamba-based RNN ensemble model to automatically perform 3-class, 4-class, and 5-class sleep staging using data from the ANNE One dual sensor using sleep stages inferred from concurrent PSG recordings. The model meets or surpasses performance metrics in related literature when applied to our ANNE One dataset, especially for wearable devices, and demonstrates robustness across a wide range of clinical characteristics. Moreover, the model's capability was achieved without prescreening and excluding large numbers of patients or epochs for reasons of signal quality. Compared to PSGs, the ANNE One sensor is far less intrusive of a device and procedure, which is significant for expanding the accessibility of sleep measurement for older and remote-living adults at greater risk of sleep-related illness.  Our model provides a scalable, accurate minimally-intrusive approach to the ambulatory assessment of sleep staging without the need for EEG.
\pagebreak
\section*{Financial Disclosure Statement}
The authors have no financial conflicts of interest to disclose. 

This research is supported by the Centre for Aging and Brain Health Innovation, Canadian Institutes of Health Research, and National Institute on Aging.

This research is not funded by Sibel Health, the manufacturer of the ANNE One device.

\section*{Non-Financial Disclosure Statement}
The authors have no non-financial conflicts of interest to disclose.

\section*{Acknowledgments}
Special thanks to Prof. Paul Kushner and Prof. Dylan Jones from the Department of Physics at the University of Toronto for offering various forms of support for this research during its infancy.

\section*{Data Availability}
All data will be made available through the Canadian Federated Research Data Repository at https://www.frdr-dfdr.ca/repo/ upon publication.  All code will be made available at https://github.com/OSHS2019/ upon publication.   

\pagebreak
\bibliographystyle{unsrtnat}
\bibliography{anne}

\pagebreak

\pagebreak
\setcounter{equation}{0}
\setcounter{figure}{0}
\setcounter{table}{0}
\makeatletter
\renewcommand{\theequation}{S\arabic{equation}}
\renewcommand{\thefigure}{S\arabic{figure}}
\renewcommand{\thetable}{S\arabic{table}}
\renewcommand{\thesection}{S}

\section{Supplementary Materials}

\subsection{Mamba Architecture}\label{sec:mamba}
\subsubsection{Selective SSM}\label{sec:sssm}
At the core of the Mamba architecture is the state space model (SSM). For some single channel input $x\in\mathbb{R}$, $N$-dimensional latent state $\textbf{h}\in\mathbb{R}^N$, single channel output $y\in\mathbb{R}$, and continuous time $t\in\mathbb{R}$, a SSM is governed by a system of first-order linear ordinary differential equations (ODE):
\begin{align}
\textbf{h}'(t) &= \bm{A}\,\textbf{h}(t)+\bm{B}\,x(t) \label{eq:cssmh}\\
y(t) &= \bm{C}\,\textbf{h}(t) + \bm{D}\,x(t)\\
\intertext{where $\bm{A}\in\mathbb{R}^{N\times N}$, $\bm{B}\in\mathbb{R}^{N\times 1}$, $\bm{C}\in\mathbb{R}^{1\times N}$, $\bm{D}\in\mathbb{R}^{1\times 1}$ are matrices that parameterize the SSM. However, in the context of sequence modeling, we require the SSM to work with discrete time steps $t\in\mathbb{Z}$. To this end we apply zero-order hold (ZOH) discretization to $\bm{A}$ and $\bm{B}$ in \eqref{eq:cssmh} such that:}
{\bm{\bar{A}}}&=\text{exp}(\Delta\bm{ A})\\
{\bm{\bar{B}}}&=(\Delta\bm{ A})^{-1}\!\bigl(\text{exp}(\Delta{\bm{ A}})-\bm{I}\bigr)\cdot\Delta\bm{ B}
\intertext{
where ${\Delta}\in\mathbb{R}^+$ is the ZOH discretization time step, which governs the magnitude of each discrete update to the system's trajectory. As such, we can reformulate the SSM from a system of ODEs to a recurrence:
}
\textbf{h}_t &= \bm{\bar{A}}\,\textbf{h}_{t-1}+\bm{\bar{B}}\,x_t\label{eq:ssmh}\\
y_t &= \bm{C}\,\textbf{h}_t + \bm{D}\,x_t \label{eq:ssmy}
\intertext{Note that Mamba's SSM still learns the parameters $\bm{ A}$ and $\bm{ B}$ rather than the discretized ${\bm{\bar{A}}}$ and ${\bm{\bar{B}}}$ directly, where ZOH discretization is applied during forward pass. The Mamba architecture also imposes an additional structural constraint that $\bm{ A}$ is diagonal.
\newline\linebreak
The SSM as formulated by \eqref{eq:ssmh} and \eqref{eq:ssmy} is linear time-invariant (LTI), which means that all parameters remain static and input-independent across time steps. This makes encoding context-awareness and adaptive behavior challenging. Mamba's selective SSM breaks LTI, building input dependence into \eqref{eq:ssmh} and \eqref{eq:ssmy} by reparameterizing $\bm{{B}}$ and $\bm{{C}}$. For a length $L$ input sequence $\textbf{x}\in\mathbb{R}^{L}$:}
\bm{{B}} &\leftarrow \text{Linear}_N(\textbf{x})\\
\bm{{C}} &\leftarrow \text{Linear}_N(\textbf{x})
\intertext{where $\text{Linear}_d(\bullet)$ is a parameterized linear projection from the input to $\mathbb{R}^d$.
\newline\linebreak
Mamba's selective SSM additionally reparameterizes $\Delta$ to make the ZOH discretization time step adaptable to the input such that it functions similarly to RNN gating mechanisms, which is when the magnitude of each hidden state update is modulated as a function of the input: 
}
\Delta &\leftarrow \text{softplus}(\Delta +  \text{Linear}_1(\textbf{x}))
\end{align}
where $\text{softplus}(\bullet) = \log(1+\text{exp}(\bullet))$.
\\\\
For a multi-channel input with $D$ channels $\bm{X} \in \mathbb{R}^{L \times D}$, the selective SSM is applied channel-wise with a total of $DN$ hidden states. Notice that by making the selective SSM not LTI, the Mamba architecture's every hidden state update is dependent on the entire history of input such that it can capture arbitrarily long-range dependencies across the entire sequence similar to the Transformer architecture.
\pagebreak
\subsubsection{Mamba Block} \label{sec:mambaBlock}
\begin{figure}[H]
  \centerline{\includegraphics[width=0.55\textwidth]{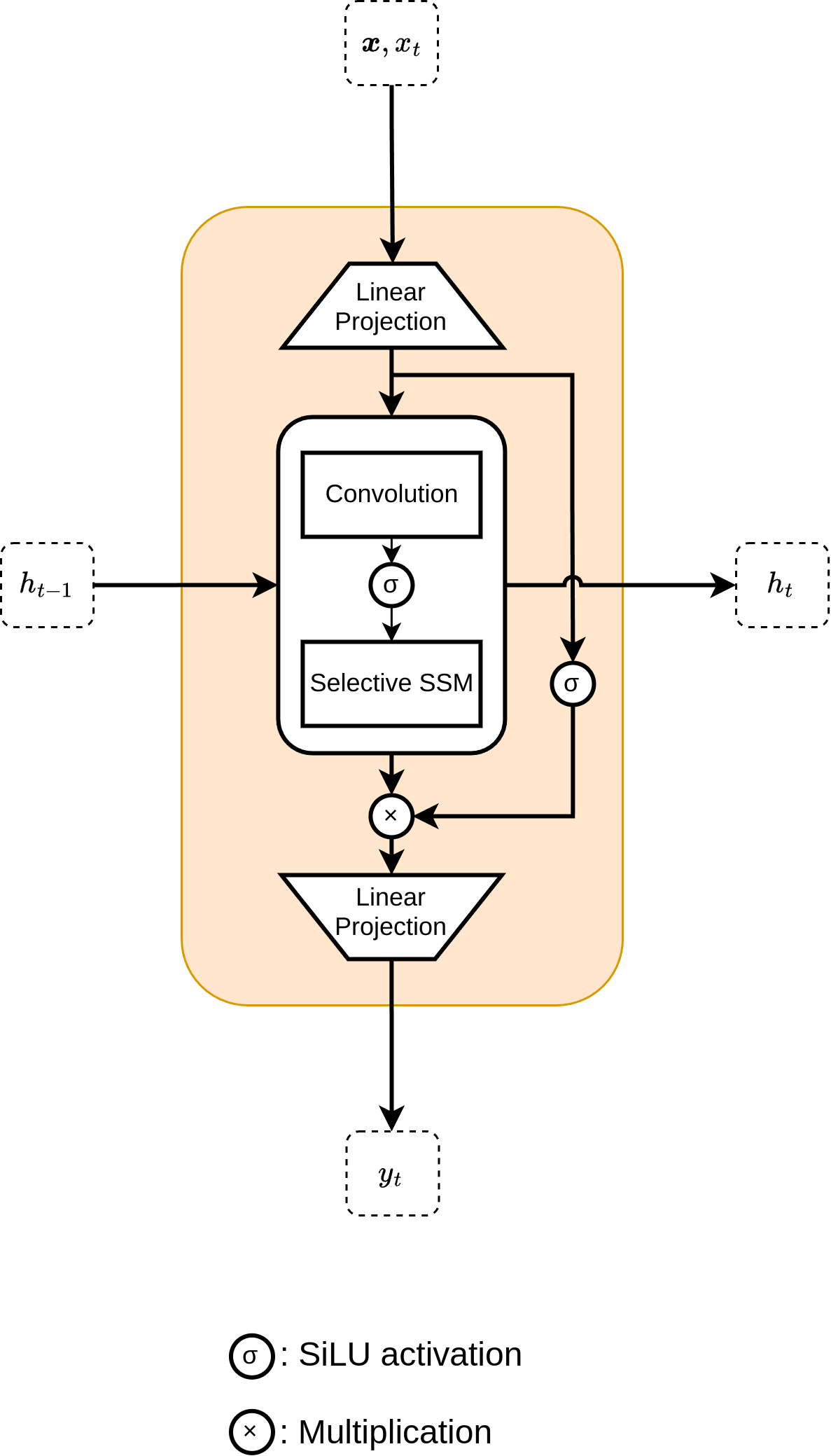}}
  \caption{Detailed architecture of the Mamba block.}
  \label{fig:mamba}
\end{figure}

A Mamba block additionally wraps the selective SSM (\ref{sec:sssm}) around linear projections, residual connections, and convolutions. The implementation of a Mamba block used in this study \cite{mambapy} is as shown on Figure \ref{fig:mamba}.
\pagebreak
\subsubsection{Bidirectional Mamba Block}
\begin{figure}[H]
  \centerline{\includegraphics[width=0.75\textwidth]{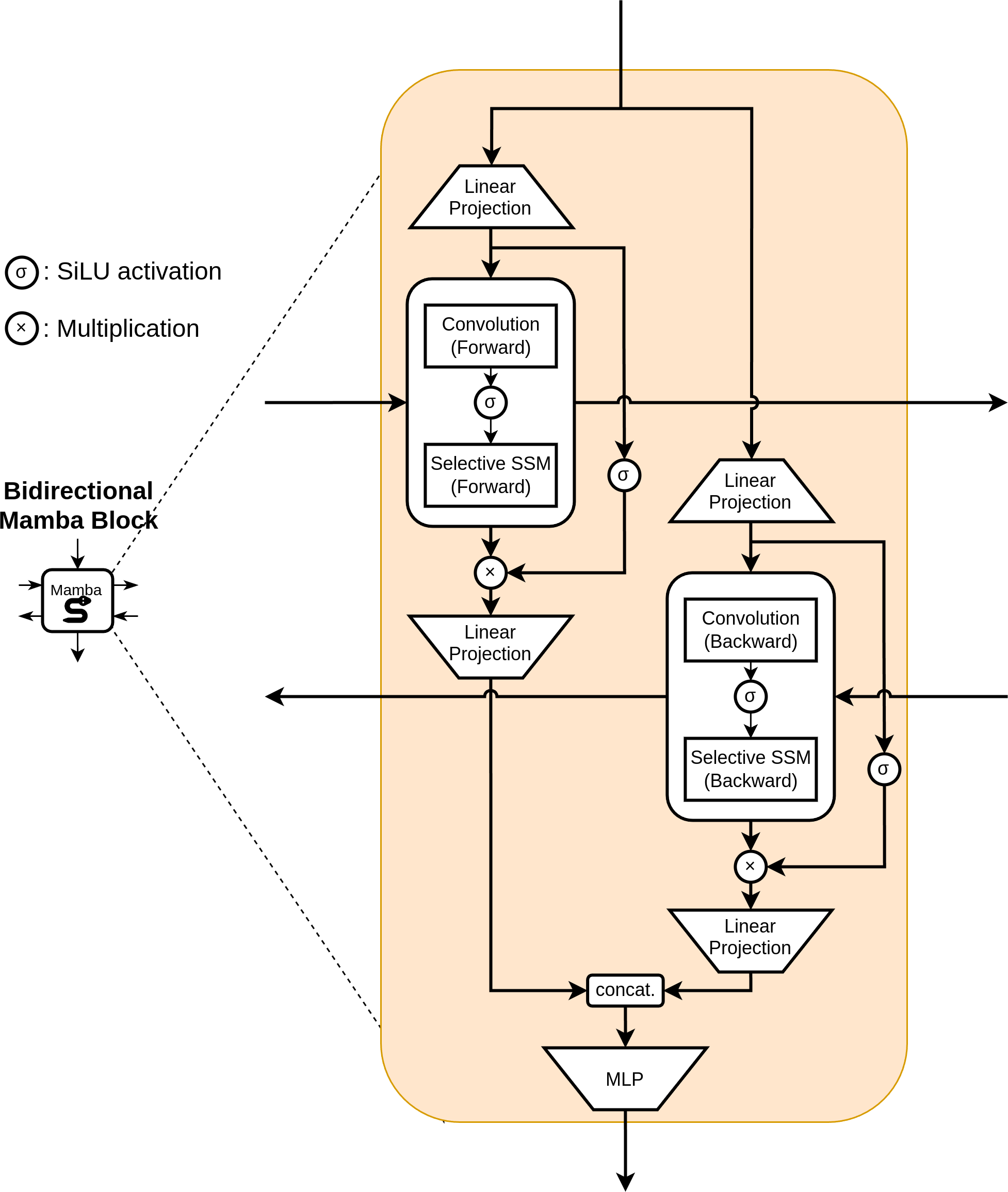}}
  \caption{Detailed architecture of the bidirectional Mamba block illustrated in Figure \ref{fig:pipelines}.}
  \label{fig:bimamba}
\end{figure}

Our bidirectional Mamba block consists of two Mamba blocks (\ref{sec:mambaBlock}). One models the sequence in chronological order, while the other models the sequence in reverse-chronological order, as shown on Figure \ref{fig:bimamba}.

\pagebreak
\subsection{Summary Statistics by Groups}\label{sec:summary-statistics-detailed}

\begin{table}[H]
\caption{Sleep variables of study participants stratified by age and sleep apnea with $t$-test $p$-values to indicate the differences' significances. Values are formatted as mean $\pm$ standard deviation.}
\centerline{\begin{tabular}{l|ccc|ccc}
\Xhline{3\arrayrulewidth}

\rowcolor[HTML]{EFEFEF} 
\textit{Variable\phantom{\Large{A}}} & \textit{Age $\geq$ 65} & \textit{Age $<$ 65} & \multicolumn{1}{c|}{\cellcolor[HTML]{EFEFEF}\textit{p}} & \textit{AHI $\geq 5$} & \textit{AHI $<$ 5} & \multicolumn{1}{c}{\cellcolor[HTML]{EFEFEF}\textit{p}} \\ \hline

Age [y]\phantom{\Large{A}} &  &  &  & 64.01 ± 14.34 & 51.82 ± 18.07 & 1.48$\times 10^{-11}$ \\
Sex [\% Female] & 44.52 ± 49.70 & 52.83 ± 49.92 & 0.123 & 40.36 ± 49.06 & 57.29 ± 49.47 & 1.34$\times 10^{-3}$ \\
AHI & 14.23 ± 16.64 & 9.90 ± 17.59 & 0.0203 &  &  &  \\
TST [h] & 4.61 ± 1.29 & 5.43 ± 1.48 & 1.46$\times 10^{-7}$ & 4.86 ± 1.35 & 5.30 ± 1.52 & 4.82$\times 10^{-3}$ \\
PLMI & 23.34 ± 35.41 & 9.22 ± 16.82 & 7.96$\times 10^{-7}$ & 16.64 ± 25.38 & 13.55 ± 28.18 & 0.281 \\
Body Mass Index & 27.89 ± 4.59 & 27.90 ± 6.19 & 0.986 & 29.45 ± 5.72 & 26.55 ± 5.13 & 7.54$\times 10^{-7}$ \\
Sleep Efficiency [\%] & 65.81 ± 16.40 & 75.84 ± 17.81 & 1.27$\times 10^{-7}$ & 69.61 ± 17.48 & 73.61 ± 18.13 & 0.0356 \\
Awake [\%] & 46.64 ± 15.00 & 37.10 ± 16.47 & 5.22$\times 10^{-8}$ & 43.01 ± 16.08 & 39.25 ± 16.78 & 0.0321 \\
N1 Sleep [\%] & 12.13 ± 6.64 & 10.43 ± 5.83 & 0.0110 & 13.15 ± 6.97 & 9.37 ± 4.88 & 5.41$\times 10^{-9}$ \\
N2 Sleep [\%] & 25.87 ± 10.31 & 30.06 ± 10.46 & 2.17$\times 10^{-4}$ & 26.77 ± 11.03 & 29.72 ± 10.02 & 8.59$\times 10^{-3}$ \\
N3 Sleep [\%] & 8.20 ± 6.73 & 12.64 ± 8.48 & 2.38$\times 10^{-7}$ & 9.00 ± 7.43 & 12.41 ± 8.34 & 6.36$\times 10^{-5}$ \\
REM Sleep [\%] & 7.16 ± 5.20 & 9.77 ± 6.40 & 5.96$\times 10^{-5}$ & 8.07 ± 5.84 & 9.25 ± 6.22 & 0.0679 \\
\Xhline{3\arrayrulewidth}
\end{tabular}}
\end{table}


    


\pagebreak

\pagebreak
\subsection{Model Architecture}\label{sec:detailed-architecture}

Below is the printed output of a $n$-class RNN model (for $n \in \{3, 4, 5\}$; for binary models, $n = 1$) with exact hyperparameter values. The selective SSM parameters are contained within \texttt{x\_proj} and \texttt{dt\_proj}. Softmax probabilities found using the output of this RNN model are used to represent the hypnodensities of the regular RNN.

\begin{small}
\begin{verbatim}
AnneNet(
  (net): MambaNet(
    (mamba_forward): Mamba(
      (layers): ModuleList(
        (0-2): 3 x ResidualBlock(
          (mixer): MambaBlock(
            (in_proj): Linear(in_features=10, out_features=40, bias=True)
            (conv1d): Conv1d(20, 20, kernel_size=(4,), stride=(1,), padding=(3,), groups=20)
            (x_proj): Linear(in_features=20, out_features=33, bias=False)
            (dt_proj): Linear(in_features=1, out_features=20, bias=True)
            (out_proj): Linear(in_features=20, out_features=10, bias=True)
          )
          (norm): RMSNorm()
        )
      )
    )
    (mamba_backward): Mamba(...same parameters as mamba_forward...)
    (pre_mlp): Sequential(
      (0): BatchNorm1d(140, eps=1e-05, momentum=0.1, affine=True, track_running_stats=True)
      (1): Linear(in_features=140, out_features=100, bias=True)
      (2): LeakyReLU(negative_slope=0.01)
      (3): BatchNorm1d(100, eps=1e-05, momentum=0.1, affine=True, track_running_stats=True)
      (4): Linear(in_features=100, out_features=100, bias=True)
      (5): LeakyReLU(negative_slope=0.01)
      (6): BatchNorm1d(100, eps=1e-05, momentum=0.1, affine=True, track_running_stats=True)
      (7): Linear(in_features=100, out_features=100, bias=True)
      (8): LeakyReLU(negative_slope=0.01)
      (9): BatchNorm1d(100, eps=1e-05, momentum=0.1, affine=True, track_running_stats=True)
      (10): Linear(in_features=100, out_features=10, bias=True)
    )
    (post_mlp): Sequential(
      (0): BatchNorm1d(20, eps=1e-05, momentum=0.1, affine=True, track_running_stats=True)
      (1): Linear(in_features=20, out_features=100, bias=True)
      (2): LeakyReLU(negative_slope=0.01)
      (3): BatchNorm1d(100, eps=1e-05, momentum=0.1, affine=True, track_running_stats=True)
      (4): Linear(in_features=100, out_features=100, bias=True)
      (5): LeakyReLU(negative_slope=0.01)
      (6): BatchNorm1d(100, eps=1e-05, momentum=0.1, affine=True, track_running_stats=True)
      (7): Linear(in_features=100, out_features=100, bias=True)
      (8): LeakyReLU(negative_slope=0.01)
      (9): BatchNorm1d(100, eps=1e-05, momentum=0.1, affine=True, track_running_stats=True)
      (10): Linear(in_features=100, out_features=n, bias=True)
    )
  )
)
\end{verbatim}
\end{small}

\pagebreak

The suite of decision tree architectures below are grouped by base model. Each configuration (separated by semicolons) is a separate model. All omitted parameters are set to the default values in Scikit-learn or XGBoost.

\begin{itemize}
\item Histogram Gradient Boosting: \texttt{default parameters; max depth = 5; learning rate = 0.05; learning rate = 0.0333; learning rate = 0.01}
\item {Random Forest:} \texttt{($n$ = 3, max depth = 7); ($n$ = 5, max depth = 5); ($n$ = 10, max depth = 3); ($n$ = 10, max depth = 3)}
\item {Extra Trees:} \texttt{($n$ = 10, max depth = 3)}
\item {AdaBoost:} \texttt{$n$ = 3; $n$ = 5}
\item {XGBoost:} \texttt{($n$ = 5, $\eta$ = 0.5); ($n$ = 10, $\eta$ = 0.5)}
\end{itemize}

The final 3-class, 4-class, and 5-class RNN ensemble models all chose a histogram gradient boosting classifier with learning rate $\alpha = 0.01$. In general, gradient boosting classifiers use many decision trees where tree $T_i$ is trained on the erroneous predictions of $T_{i-1}$, which gradually refines performance as $i$ increases. For multi-class classification, regression trees compute logits for each class before applying softmax:

\vspace{-2mm}

$$\text{logit}(x) = T_1(x) + \alpha \big(T_2(x) + T_3(x) + ...\big)$$

As $\alpha$ is miniscule, the RNN ensemble's logits appear largely dominated by the initial prediction from the baseline decision tree $T_1$. The first tree makes simple and underfitted predictions, resulting in stable but highly-uncertain probabilities such as those in Figure \ref{fig:monolith}.

\pagebreak
\subsection{Additional Model Visualizations}\label{sec:ensemble-mode}

\begin{figure}[H]
\centerline{\includegraphics[width=1.15\textwidth]{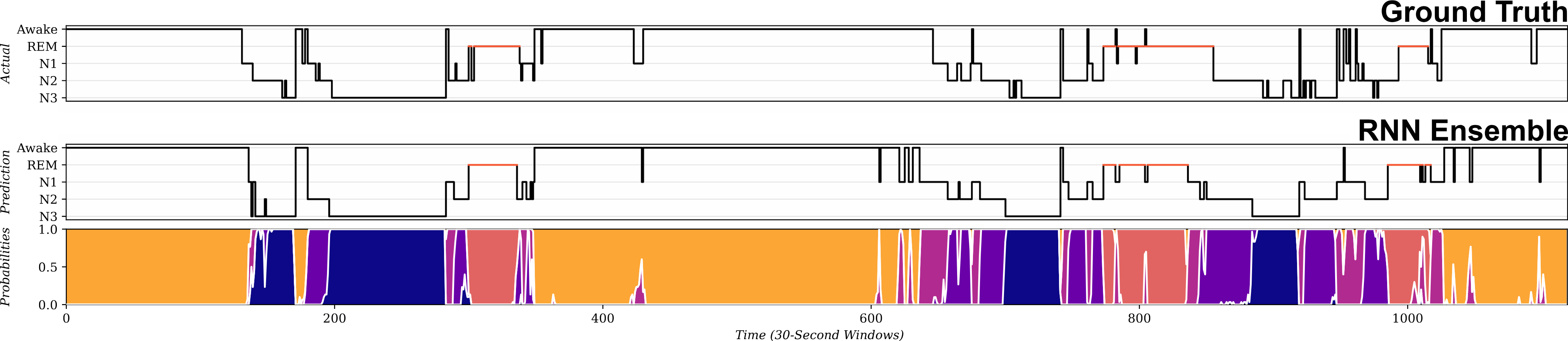}}
  \caption{Predicted sleep stages (hypnograms) and class probabilities (hypnodensities) for a typical single full-night recording in the test set for the 5-class RNN ensemble model against the ground truth hypnogram. The predicted hypnodensity is calculated using an alternate method to Figure \ref{fig:monolith} that groups 1-second predictions into 30-second buckets, interpreting predictions in each bucket as a probability distribution, and taking the most common sleep stage in the bucket.}
  \label{fig:alternate-hypnodensity}
\end{figure}

\begin{figure}[H]
\centerline{
    \includegraphics[width=0.575\textwidth]{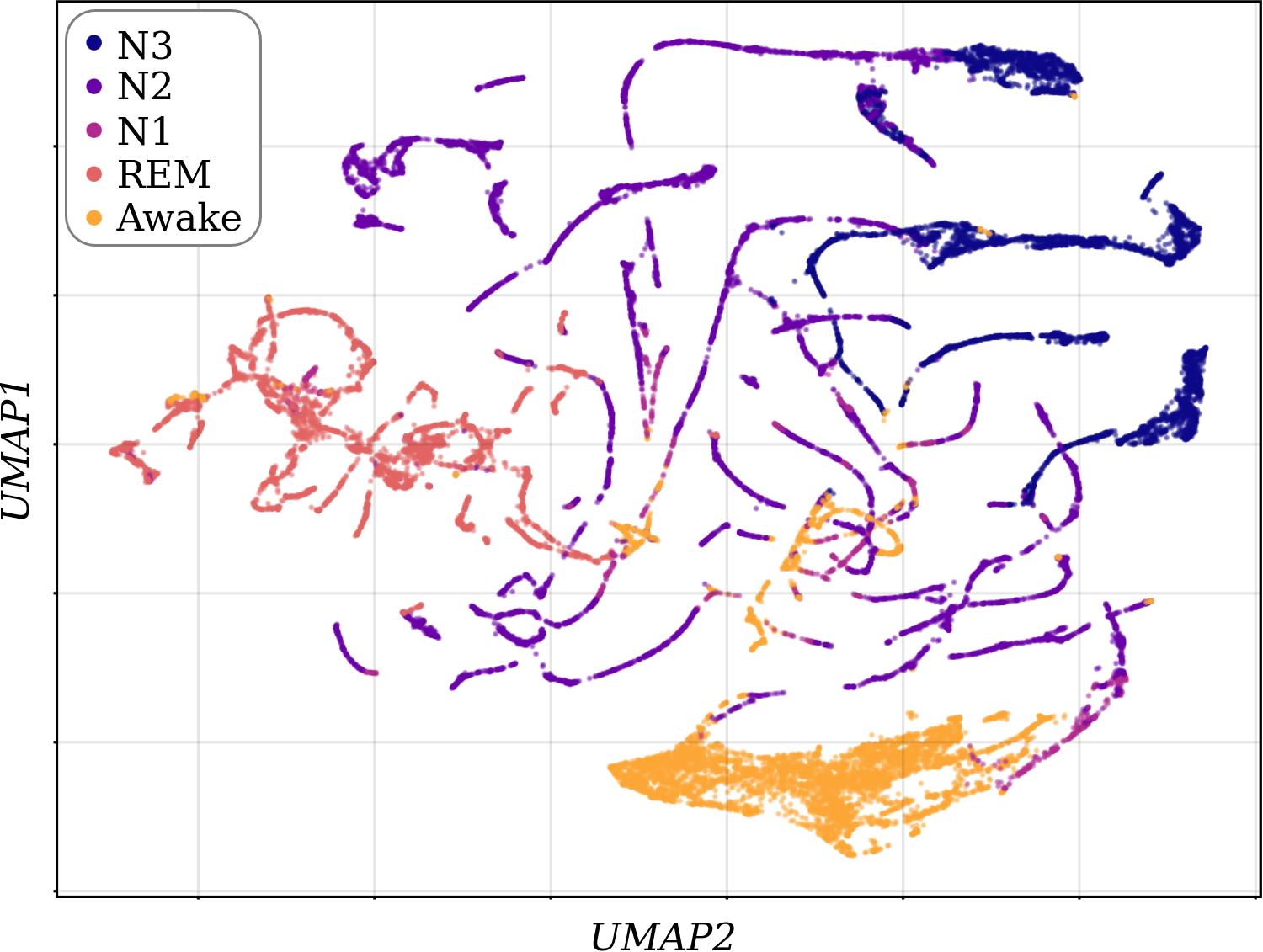}
    \includegraphics[width=0.575\textwidth]{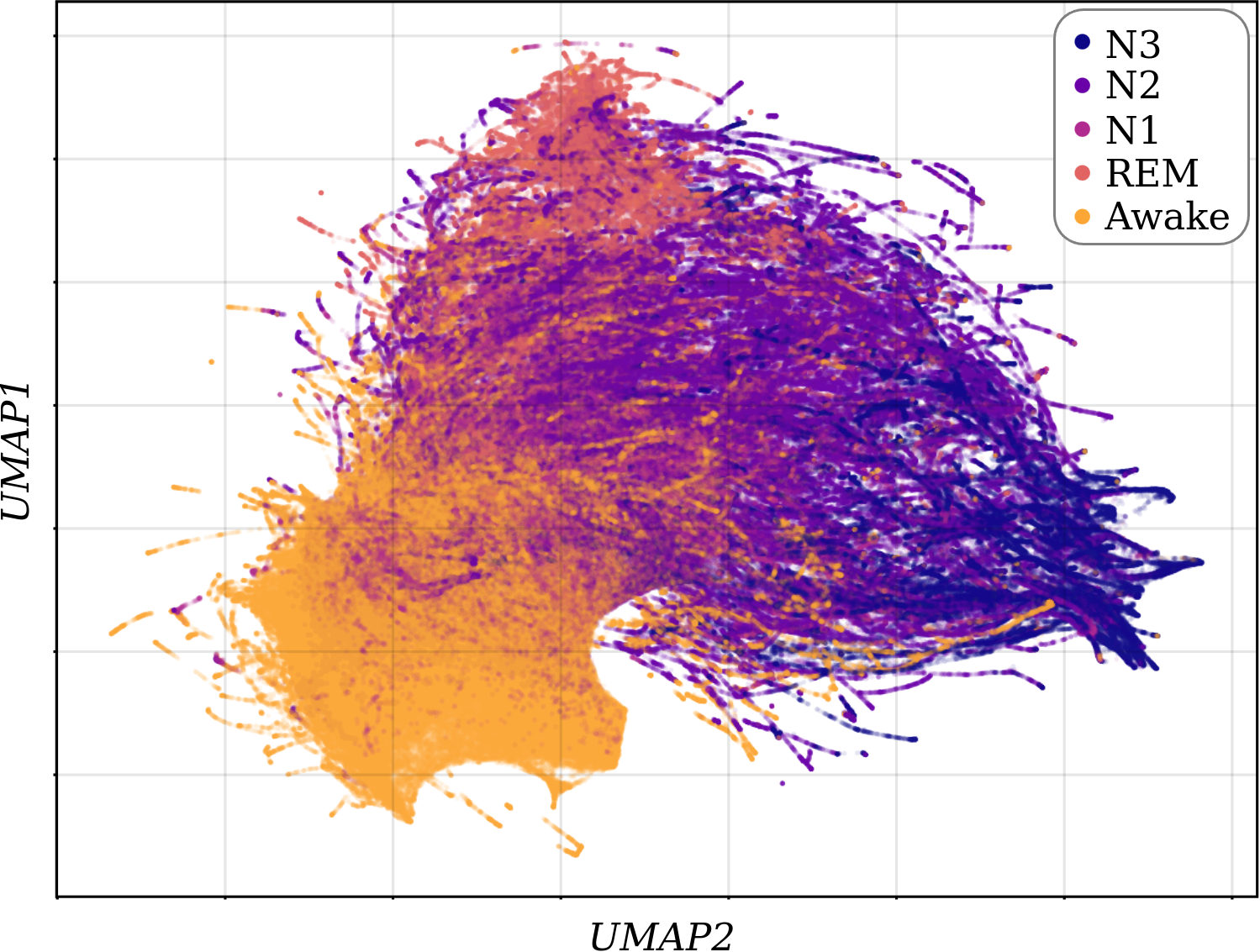}
}
  \caption{UMAP projections of concatenated bidirectional Mamba hidden state vectors from the 5-class RNN model for a single full-night recording for an individual healthy patient (left) and all patients (right) in the test set, colored by ground truth sleep stage.}
  \label{fig:umap}
\end{figure}
UMAP projections on Figure \ref{fig:umap} are generated using the following hyperparameters:

\begin{multicols}{2}

\begin{itemize}
  \item \texttt{metric}, \texttt{output\_metric}: euclidean
  \item \texttt{n\_neighbors}: 15
  \item \texttt{n\_components}: 2
  \item \texttt{learning\_rate}: 1.0
  \item \texttt{min\_dist}: 0.1
  \item \texttt{spread}: 1.0
  \item \texttt{set\_op\_mix\_ratio}: 1.0
\end{itemize}

\columnbreak

\begin{itemize}
    \item \texttt{local\_connectivity}: 1
  \item \texttt{negative\_sample\_rate}: 5
  \item \texttt{target\_n\_neighbors}: -1
  \item \texttt{target\_weight}: 0.5
  \item \texttt{dens\_lambda}: 2.0
  \item \texttt{dens\_frac}: 0.3
  \item \texttt{dens\_var\_shift}: 0.1
\end{itemize}

\end{multicols}

\pagebreak
\subsection{Class-Wise Evaluation Metrics}\label{sec:classwise-metrics}

\begin{table}[H]
\caption{Class-wise metrics for all $n$-class models on the test set, calculated by the metrics as applied to binary labels indicating the presence or absence of a given class. The best-performing model for each metric is bolded.}
\label{tab:results}
\footnotesize
  \setlength{\aboverulesep}{0pt}
  \setlength{\belowrulesep}{0pt}
  \setlength{\tabcolsep}{3.5pt}
\centerline{\begin{tabular}{r!{\vrule width 1.2pt}
>{\columncolor[HTML]{EFEFEF}}c 
>{\columncolor[HTML]{EFEFEF}}c |
>{\columncolor[HTML]{FFFFFF}}c 
>{\columncolor[HTML]{FFFFFF}}c |
>{\columncolor[HTML]{EFEFEF}}c 
>{\columncolor[HTML]{EFEFEF}}c }
\Xhline{3\arrayrulewidth}
 & \multicolumn{2}{c|}{\cellcolor[HTML]{EFEFEF}\phantom{\Large{A}}\textit{5-Class Sleep Staging}\phantom{\Large{A}}} & \multicolumn{2}{c|}{\cellcolor[HTML]{FFFFFF}\phantom{\Large{A}}\textit{4-Class Sleep Staging}\phantom{\Large{A}}} & \multicolumn{2}{c}{\cellcolor[HTML]{EFEFEF}\phantom{\Large{A}}\textit{3-Class Sleep Staging}\phantom{\Large{A}}} \\
\cline{2-7} 
\multirow{-2}{*}{\textit{Metric (\%)}} & {\cellcolor[HTML]{EFEFEF}\phantom{\Large{A}}Regular\phantom{\Large{A}}} & {\cellcolor[HTML]{EFEFEF}Ensemble} & {\cellcolor[HTML]{FFFFFF}\phantom{\Large{A}}Regular\phantom{\Large{A}}} & {\cellcolor[HTML]{FFFFFF}Ensemble} & {\cellcolor[HTML]{EFEFEF}\phantom{\Large{A}}Regular\phantom{\Large{A}}} & {\cellcolor[HTML]{EFEFEF}Ensemble} \\ 
\Xhline{3\arrayrulewidth}

 & \multicolumn{2}{c|}{\cellcolor[HTML]{EFEFEF}\textbf{\phantom{\Large{A}}Wake\phantom{\Large{A}}}} & \multicolumn{2}{c|}{\cellcolor[HTML]{FFFFFF}\textbf{\phantom{\Large{A}}Wake\phantom{\Large{A}}}} & \multicolumn{2}{c}{\cellcolor[HTML]{EFEFEF}\textbf{\phantom{\Large{A}}Wake\phantom{\Large{A}}}} \\
Balanced Accuracy & \textbf{86.47} & 86.19  & \textbf{87.93} & 87.87  & 87.97 & \textbf{88.27} \\
Precision & 87.64 & \textbf{88.44}  & 88.34 & \textbf{88.67}  & 88.24 & \textbf{88.82} \\
Recall & 87.64 & \textbf{88.04}  & 88.32 & \textbf{88.70}  & 88.12 & \textbf{88.85} \\
F1 & 87.54 & \textbf{87.81}  & 88.33 & \textbf{88.66}  & 88.16 & \textbf{88.83} \\
Cohen's $\kappa$ & 73.93 & \textbf{74.41}  & 75.74 & \textbf{76.32}  & \textbf{75.45} & 72.89 \\
MCC & 74.11 & \textbf{75.17}  & 75.74 & \textbf{76.37}  & 75.49 & \textbf{76.74} \\ \hline
 & \multicolumn{2}{c|}{\cellcolor[HTML]{EFEFEF}\textbf{\phantom{\Large{A}}N1 Sleep\phantom{\Large{A}}}} & \multicolumn{2}{c|}{\cellcolor[HTML]{FFFFFF}\textbf{\phantom{\Large{A}}N1/N2 Sleep\phantom{\Large{A}}}} & \multicolumn{2}{c}{\cellcolor[HTML]{EFEFEF}\textbf{\phantom{\Large{A}}Non-REM Sleep\phantom{\Large{A}}}} \\
Balanced Accuracy & 67.99 & \textbf{69.62}  & 71.58 & \textbf{73.47}  & 81.03 & \textbf{85.42} \\
Precision & 86.74 & \textbf{87.09} & 75.37 & \textbf{75.61}  & 81.90 & \textbf{85.50} \\
Recall  & \textbf{81.74} & 79.31  & 75.06 & \textbf{75.81}   & 80.99 & \textbf{85.41} \\
F1 & \textbf{83.77} & 82.25   & 74.00 & \textbf{75.36}  & 80.87 & \textbf{85.41} \\
Cohen's $\kappa$ & \textbf{27.21} & 26.46   & 45.39 & \textbf{48.25}  & 62.01 & \textbf{70.83} \\
MCC & 28.61 & \textbf{29.06}& 47.03 & \textbf{48.74}  & 62.90 & \textbf{70.92} \\ \hline
 & \multicolumn{2}{c|}{\cellcolor[HTML]{EFEFEF}\textbf{\phantom{\Large{A}}N2 Sleep\phantom{\Large{A}}}} &  &  &  &    \\
Balanced Accuracy & 65.88 & \textbf{67.72} &   &  &  &  \\
Precision  & 76.25 & \textbf{76.33} &   &  &  &  \\
Recall  & 77.22 & \textbf{77.50} &  &  &  &  \\
F1  & 74.80 & \textbf{75.82} &  &    &  &  \\
Cohen's $\kappa$ & 36.71 & \textbf{39.58} &  &   &  &  \\
MCC & 39.68 & \textbf{41.29} &  &   &  &  \\ \hline
 & \multicolumn{2}{c|}{\cellcolor[HTML]{EFEFEF}\textbf{\phantom{\Large{A}}N3 Sleep\phantom{\Large{A}}}} & \multicolumn{2}{c|}{\cellcolor[HTML]{FFFFFF}\textbf{\phantom{\Large{A}}N3 Sleep\phantom{\Large{A}}}} &  &   \\
Balanced Accuracy& 79.40 & \textbf{81.16}  & 80.08 & \textbf{82.18} &  &  \\
Precision & 90.86 & \textbf{91.71}   & 90.92 & \textbf{91.72} &  &    \\
Recall  & 87.60 & \textbf{89.66} & 86.85 & \textbf{88.59} &  &   \\
F1  & 88.82 & \textbf{90.45}   & 88.32 & \textbf{89.72} &  &   \\
Cohen's $\kappa$  & 46.36 & \textbf{52.35}   & 45.47 & \textbf{50.70} &  &  \\
MCC & 48.08 & \textbf{53.44}  & 47.77 & \textbf{52.60} &  &  \\ \hline
 & \multicolumn{2}{c|}{\cellcolor[HTML]{EFEFEF}\textbf{\phantom{\Large{A}}REM Sleep\phantom{\Large{A}}}} & \multicolumn{2}{c|}{\cellcolor[HTML]{FFFFFF}\textbf{\phantom{\Large{A}}REM Sleep\phantom{\Large{A}}}} & \multicolumn{2}{c}{\cellcolor[HTML]{EFEFEF}\textbf{\phantom{\Large{A}}REM Sleep\phantom{\Large{A}}}} \\
Balanced Accuracy  & 82.31 & \textbf{86.60} & 82.99 & \textbf{88.67}  & 84.13 & \textbf{89.26} \\
Precision & 92.70 & \textbf{94.69} & 92.63 & \textbf{94.97}  & 92.52 & \textbf{94.87} \\
Recall& 91.39 & \textbf{94.31}  & 90.70 & \textbf{94.26}   & 89.04 & \textbf{93.81} \\
F1 & 91.91 & \textbf{94.46} & 91.42 & \textbf{94.52}  & 90.25 & \textbf{94.18} \\
Cohen's $\kappa$  & 56.59 & \textbf{69.06}   & 55.18 & \textbf{70.06}   & 51.95 & \textbf{68.78} \\
MCC & 57.27 & \textbf{69.23}  & 56.35 & \textbf{70.54}  & 54.38 & \textbf{69.61} \\ \Xhline{3\arrayrulewidth}
\end{tabular}}
\end{table}

\pagebreak
\subsection{CRNN Model Architecture}\label{sec:crnn-info}

\begin{figure}[H]
  \centerline{\includegraphics[width=\textwidth]{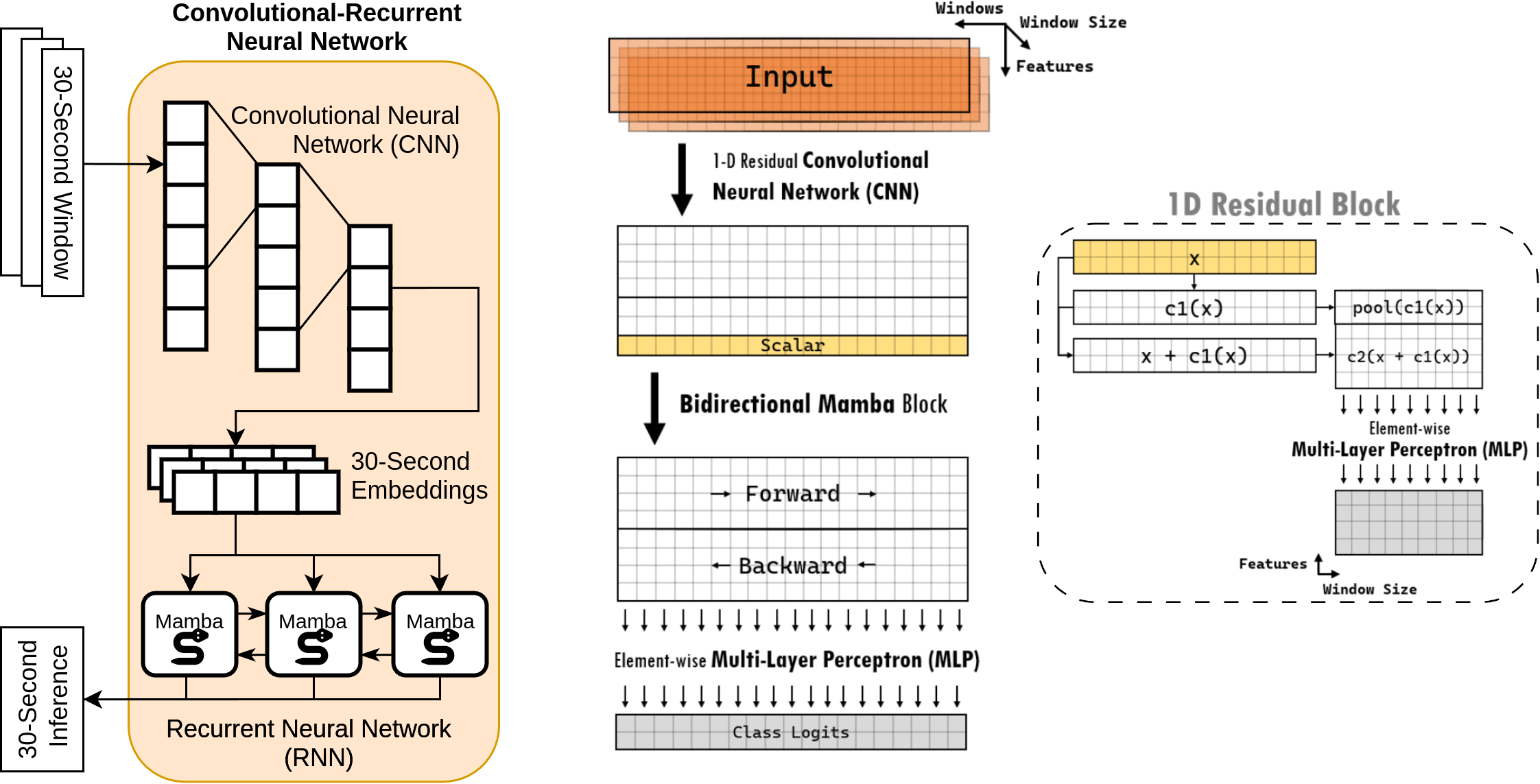}}
  \caption{Overview and implementation details of the CRNN architecture. Each 1D residual block increases the number of features and halves the window size.}
  \label{fig:crnn}
\end{figure}

The CRNN model reorganizes the input data into 30-second windows mirroring the 30-second windows used in visual annotation of PSG data according to AASM criteria. As shown in Figure \ref{fig:crnn}, the CRNN model comprises:
\begin{enumerate}[label=\arabic*)]
 \item A convolutional component, consisting of four 1D residual blocks (see \ref{fig:crnn}, right) which continually extract features and downsize the window size axis, flattening it into an embedding vector per each 30-second window.
 \item A recurrent component, consisting of a bidirectional (same as RNN model, \ref{sec:mamba}) three-layer Mamba block that processes the embedding vectors sequentially, followed by an element-wise multi-layer perceptron (MLP) that diminishes the feature-size axis into logits for each class.
\end{enumerate}

\pagebreak
Below is the printed output of the layers of an $n$-class CRNN model (for $n \in \{3, 4, 5\}$; for binary models, $n=1$) with exact hyperparameter values. 

\begin{small}
\begin{verbatim}
CRNN(
  (conv_blocks): ModuleList(
    (0): ResidualBlock(
      (net1): Sequential(
        (0): BatchNorm1d(42, eps=1e-05, momentum=0.1, affine=True, track_running_stats=True)
        (1): Conv1d(42, 42, kernel_size=(7,), stride=(1,), padding=(3,))
        (2): LeakyReLU(negative_slope=0.01)
      )
      (net2): Sequential(
        (0): BatchNorm1d(42, eps=1e-05, momentum=0.1, affine=True, track_running_stats=True)
        (1): Conv1d(42, 64, kernel_size=(7,), stride=(2,), padding=(3,))
        (2): LeakyReLU(negative_slope=0.01)
      )
      (mlp): Sequential(
        (0): Linear(in_features=106, out_features=64, bias=True)
        (1): LeakyReLU(negative_slope=0.01)
        (2): Linear(in_features=64, out_features=64, bias=True)
        (3): LeakyReLU(negative_slope=0.01)
        (4): Linear(in_features=64, out_features=64, bias=True)
        (5): LeakyReLU(negative_slope=0.01)
      )
      (pool): MaxPool1d(kernel_size=2, stride=2, padding=0, dilation=1, ceil_mode=False)
    )
    (1): ResidualBlock(
      (net1): Sequential(
        (0): BatchNorm1d(64, eps=1e-05, momentum=0.1, affine=True, track_running_stats=True)
        (1): Conv1d(64, 64, kernel_size=(7,), stride=(1,), padding=(3,))
        (2): LeakyReLU(negative_slope=0.01)
      )
      (net2): Sequential(
        (0): BatchNorm1d(64, eps=1e-05, momentum=0.1, affine=True, track_running_stats=True)
        (1): Conv1d(64, 84, kernel_size=(7,), stride=(2,), padding=(3,))
        (2): LeakyReLU(negative_slope=0.01)
      )
      (mlp): Sequential(
        (0): Linear(in_features=148, out_features=84, bias=True)
        (1): LeakyReLU(negative_slope=0.01)
        (2): Linear(in_features=84, out_features=84, bias=True)
        (3): LeakyReLU(negative_slope=0.01)
        (4): Linear(in_features=84, out_features=84, bias=True)
        (5): LeakyReLU(negative_slope=0.01)
      )
      (pool): MaxPool1d(kernel_size=2, stride=2, padding=0, dilation=1, ceil_mode=False)
    )
    (2): ResidualBlock(
      (net1): Sequential(
        (0): BatchNorm1d(84, eps=1e-05, momentum=0.1, affine=True, track_running_stats=True)
        (1): Conv1d(84, 84, kernel_size=(7,), stride=(1,), padding=(3,))
        (2): LeakyReLU(negative_slope=0.01)
      )
      (net2): Sequential(
        (0): BatchNorm1d(84, eps=1e-05, momentum=0.1, affine=True, track_running_stats=True)
        (1): Conv1d(84, 104, kernel_size=(7,), stride=(2,), padding=(3,))
        (2): LeakyReLU(negative_slope=0.01)
      )
      (mlp): Sequential(
        (0): Linear(in_features=188, out_features=104, bias=True)
        (1): LeakyReLU(negative_slope=0.01)
        (2): Linear(in_features=104, out_features=104, bias=True)
        (3): LeakyReLU(negative_slope=0.01)
        (4): Linear(in_features=104, out_features=104, bias=True)
        (5): LeakyReLU(negative_slope=0.01)
      )
      (pool): MaxPool1d(kernel_size=2, stride=2, padding=0, dilation=1, ceil_mode=False)
    )
    (3): ResidualBlock(
      (net1): Sequential(
        (0): BatchNorm1d(104, eps=1e-05, momentum=0.1, affine=True, track_running_stats=True)
        (1): Conv1d(104, 104, kernel_size=(7,), stride=(1,), padding=(3,))
        (2): LeakyReLU(negative_slope=0.01)
      )
      (net2): Sequential(
        (0): BatchNorm1d(104, eps=1e-05, momentum=0.1, affine=True, track_running_stats=True)
        (1): Conv1d(104, 126, kernel_size=(7,), stride=(2,), padding=(3,))
        (2): LeakyReLU(negative_slope=0.01)
      )
      (mlp): Sequential(
        (0): Linear(in_features=230, out_features=126, bias=True)
        (1): LeakyReLU(negative_slope=0.01)
        (2): Linear(in_features=126, out_features=126, bias=True)
        (3): LeakyReLU(negative_slope=0.01)
        (4): Linear(in_features=126, out_features=126, bias=True)
        (5): LeakyReLU(negative_slope=0.01)
      )
      (pool): MaxPool1d(kernel_size=2, stride=2, padding=0, dilation=1, ceil_mode=False)
    )
  )
  (rnn): MambaNet(
    (mamba_forward): Mamba(
      (layers): ModuleList(
        (0-2): 3 x ResidualBlock(
          (mixer): MambaBlock(
            (in_proj): Linear(in_features=126, out_features=504, bias=True)
            (conv1d): Conv1d(252, 252, kernel_size=(4,), stride=(1,), padding=(3,), groups=252)
            (x_proj): Linear(in_features=252, out_features=40, bias=False)
            (dt_proj): Linear(in_features=8, out_features=252, bias=True)
            (out_proj): Linear(in_features=252, out_features=126, bias=True)
          )
          (norm): RMSNorm()
        )
      )
    )
    (mamba_backward): Mamba(...same parameters as mamba_forward...)
    (mlp): Sequential(
      (0): BatchNorm1d(252, eps=1e-05, momentum=0.1, affine=True, track_running_stats=True)
      (1): Linear(in_features=252, out_features=100, bias=True)
      (2): LeakyReLU(negative_slope=0.01)
      (3): BatchNorm1d(100, eps=1e-05, momentum=0.1, affine=True, track_running_stats=True)
      (4): Linear(in_features=100, out_features=100, bias=True)
      (5): LeakyReLU(negative_slope=0.01)
      (6): BatchNorm1d(100, eps=1e-05, momentum=0.1, affine=True, track_running_stats=True)
      (7): Linear(in_features=100, out_features=100, bias=True)
    )
  )
  (rnn_activation): LeakyReLU(negative_slope=0.01)
  (mlp): Sequential(
    (0): Linear(in_features=226, out_features=64, bias=True)
    (1): LeakyReLU(negative_slope=0.01)
    (2): Linear(in_features=64, out_features=64, bias=True)
    (3): LeakyReLU(negative_slope=0.01)
    (4): Linear(in_features=64, out_features=16, bias=True)
    (5): LeakyReLU(negative_slope=0.01)
    (6): Linear(in_features=16, out_features=n, bias=True)
  )
)
\end{verbatim}
\end{small}

\pagebreak
\subsection{Additional Model Performance Plots}\label{sec:additional-metrics}

\begin{figure}[H]
\centerline{\includegraphics[width=1.15\textwidth]{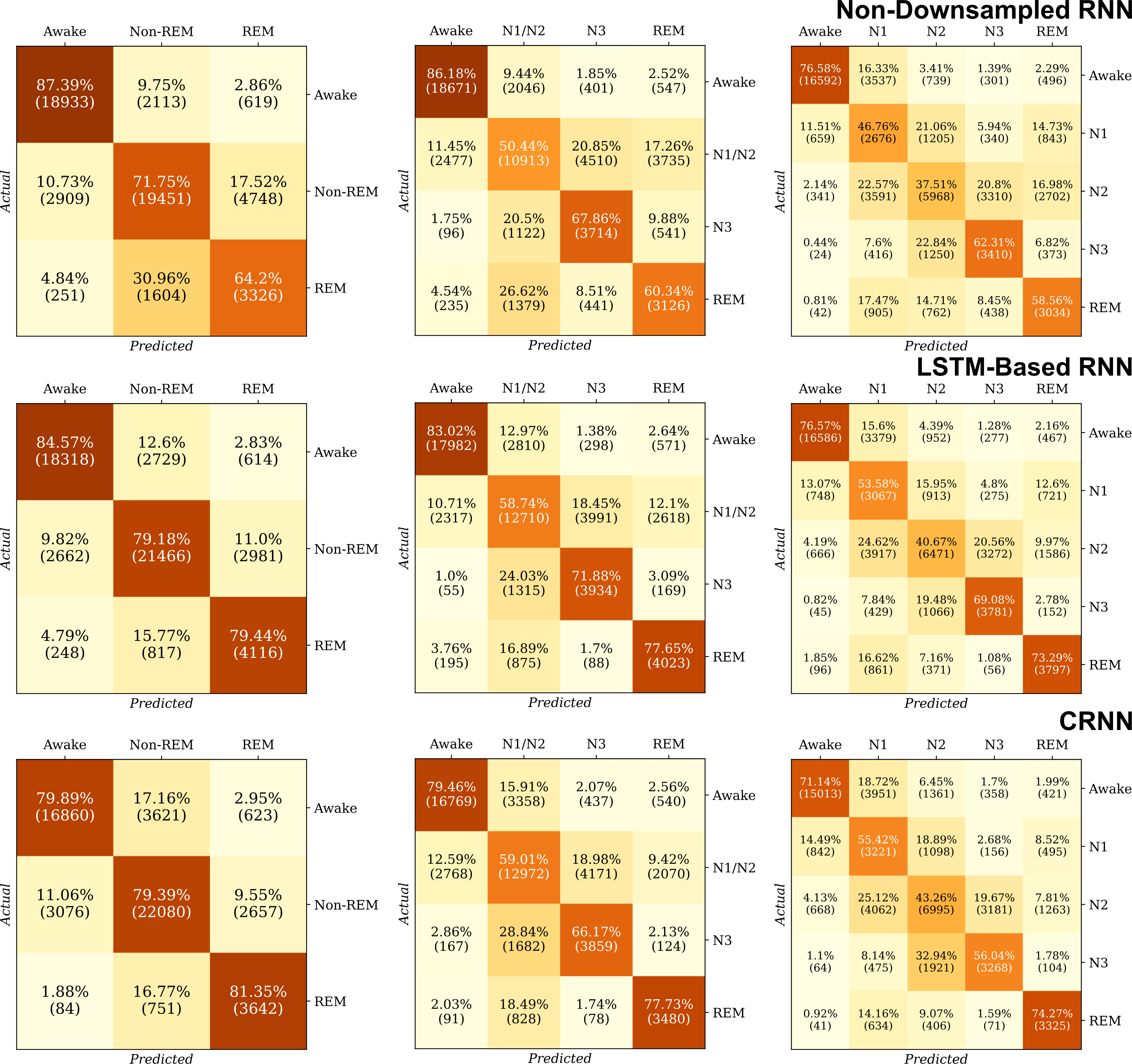}}
  \caption{Test set confusion matrices of various model ensembles for 3, 4, and 5 class sleep staging including the RNN model trained on signals that were not downsampled (top), a variant of the RNN model with the bidirectional Mamba replaced with a bidirectional LSTM (middle), and a convolutional-recurrent neural network (CRNN) model (bottom). All variants exhibit degraded performance compared to the original RNN ensemble model.}
  \label{fig:extra-conf-matrices}
\end{figure}

\begin{figure}[H]
\centerline{\includegraphics[width=1.15\textwidth]{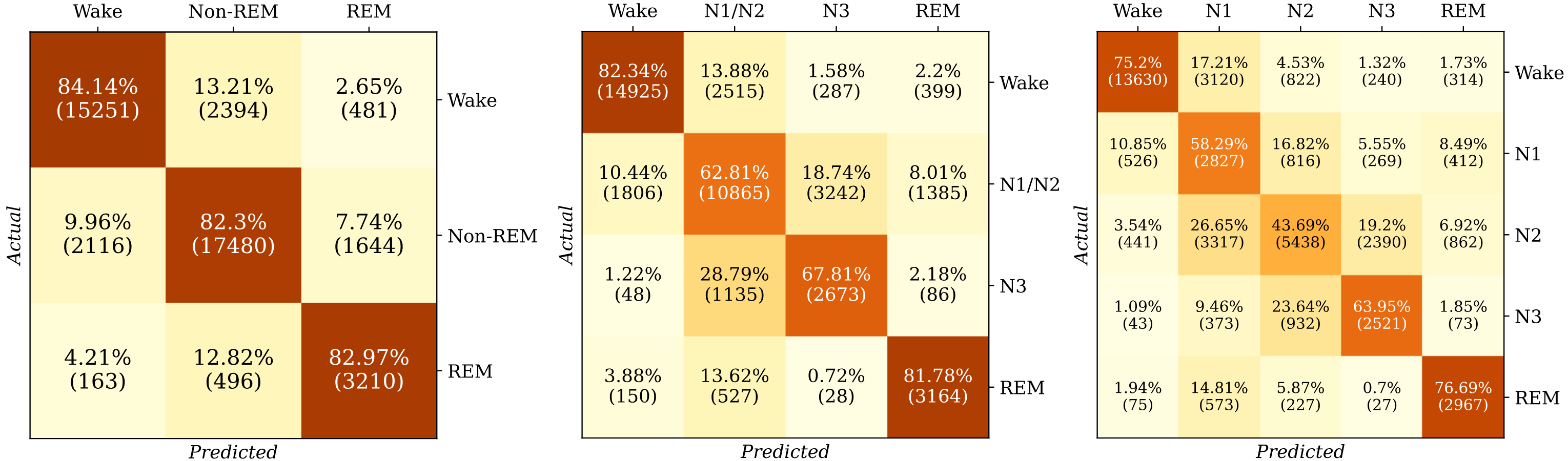}}
\end{figure}

\vspace{-6mm}

\begin{table}[H]
\footnotesize
\centerline{\begin{tabular}{c|cccccc}
\Xhline{3\arrayrulewidth} 
\rowcolor[HTML]{EFEFEF} 
\cellcolor[HTML]{EFEFEF} & \multicolumn{6}{c}{\cellcolor[HTML]{EFEFEF}\phantom{\Large{A}}\textit{Metric (\%)}\phantom{\Large{A}}} \\

\rowcolor[HTML]{EFEFEF} 
\multirow{-2}{*}{\cellcolor[HTML]{EFEFEF}\textit{Classes}} & Balanced Accuracy & Precision & Recall & F1 & Cohen's $\kappa$ & MCC \\ \Xhline{3\arrayrulewidth} 

\phantom{\Large{A}}5\phantom{\Large{A}} & 63.56 & 70.91 & 63.34 & 65.30 & 51.43 & 52.60 \\

4 & 73.68 & 75.40 & 73.15 & 73.73 & 60.11 & 60.46 \\

3 & 83.13 & 84.02 & 83.13 & 83.38 & 71.28 & 71.41 \\ \Xhline{3\arrayrulewidth} 

\end{tabular}}
\end{table}

\vspace{-6mm}

\begin{figure}[H]
\caption{Confusion matrices and macro-evaluation metrics of 3-class, 4-class, and 5-class RNN ensemble models after filtering out the healthy subset of the test set ($n$ = 43), defined as age $\geq$ 40 or AHI $\geq$ 5 or PLMI $\geq$ 5.}
  \label{fig:unhealthy-subset}
\end{figure}

\begin{figure}[H]
\centerline{\includegraphics[width=1.15\textwidth]{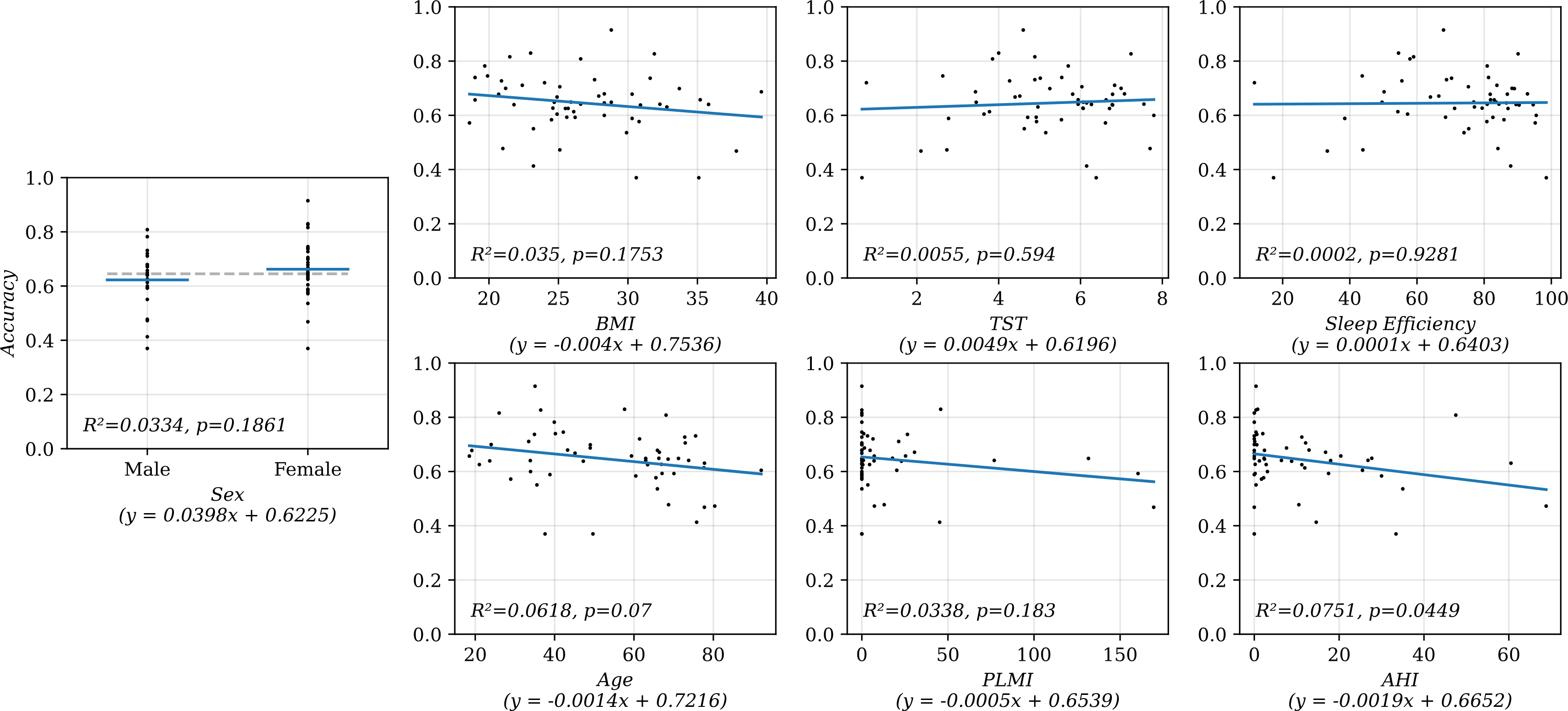}}
  \caption{Scatter plots of the 5-class RNN ensemble's recording-wise accuracy against clinical metrics comprising age, sex, PLMI, AHI, BMI, TST, and sleep efficiency for recordings in the test set. A linear regression line (or mean line in grey with group mean lines in blue) is plotted, with its equation, $\text{R}^2$, and $p$-value of slope provided below.}
  \label{fig:continuous-plots}
\end{figure}

\begin{figure}[H]
\centerline{\includegraphics[width=1.15\textwidth]{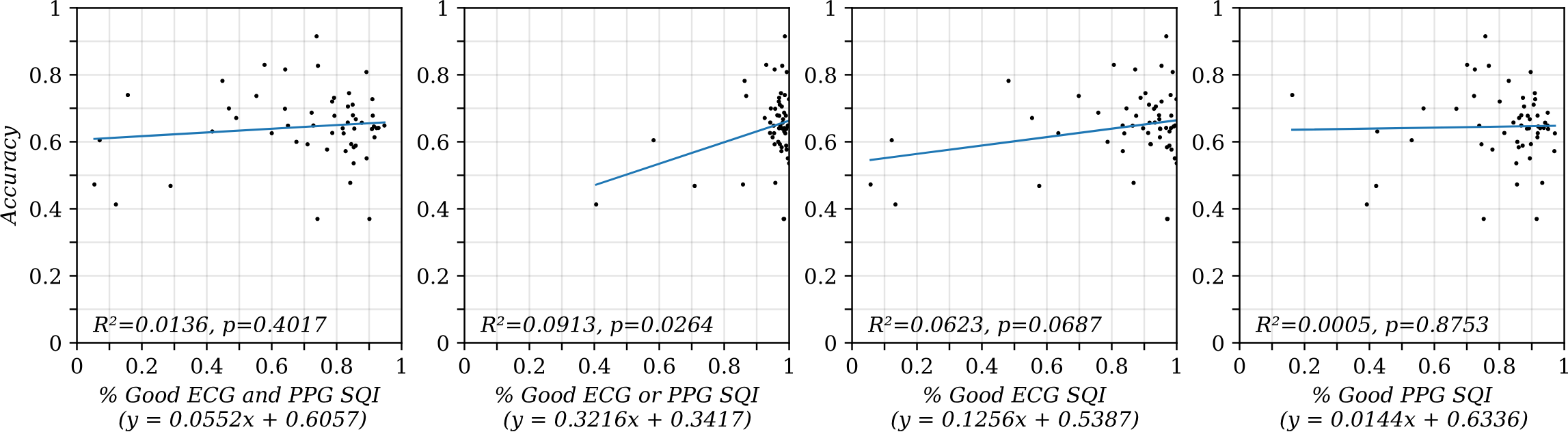}}
  \caption{Scatter plots of the 5-class RNN ensemble's recording-wise accuracy against the proportion of the recording with high SQI for ECG, PPG, either, or both in the test set. A linear regression line is plotted, with its equation, $\text{R}^2$, and $p$-value of slope provided below.}
  \label{fig:sqi-plot}
\end{figure}

\begin{figure}[H]
\centerline{\includegraphics[width=1.15\textwidth]{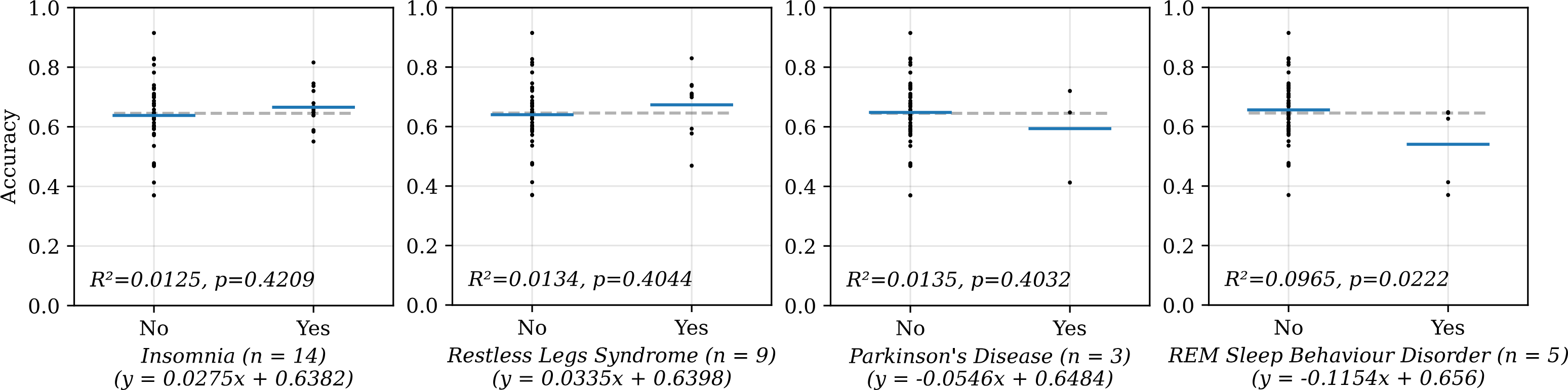}}
  \caption{Scatter plots of the 5-class RNN ensemble's recording-wise accuracy against the presence of a sleep disorder in the test set. Mean lines is plotted in grey with group mean lines in blue, with a linear regression equation, $\text{R}^2$, and $p$-value of slope provided below.}
  \label{fig:disorders}
\end{figure}

\begin{figure}[H]
\centerline{\includegraphics[width=\textwidth]{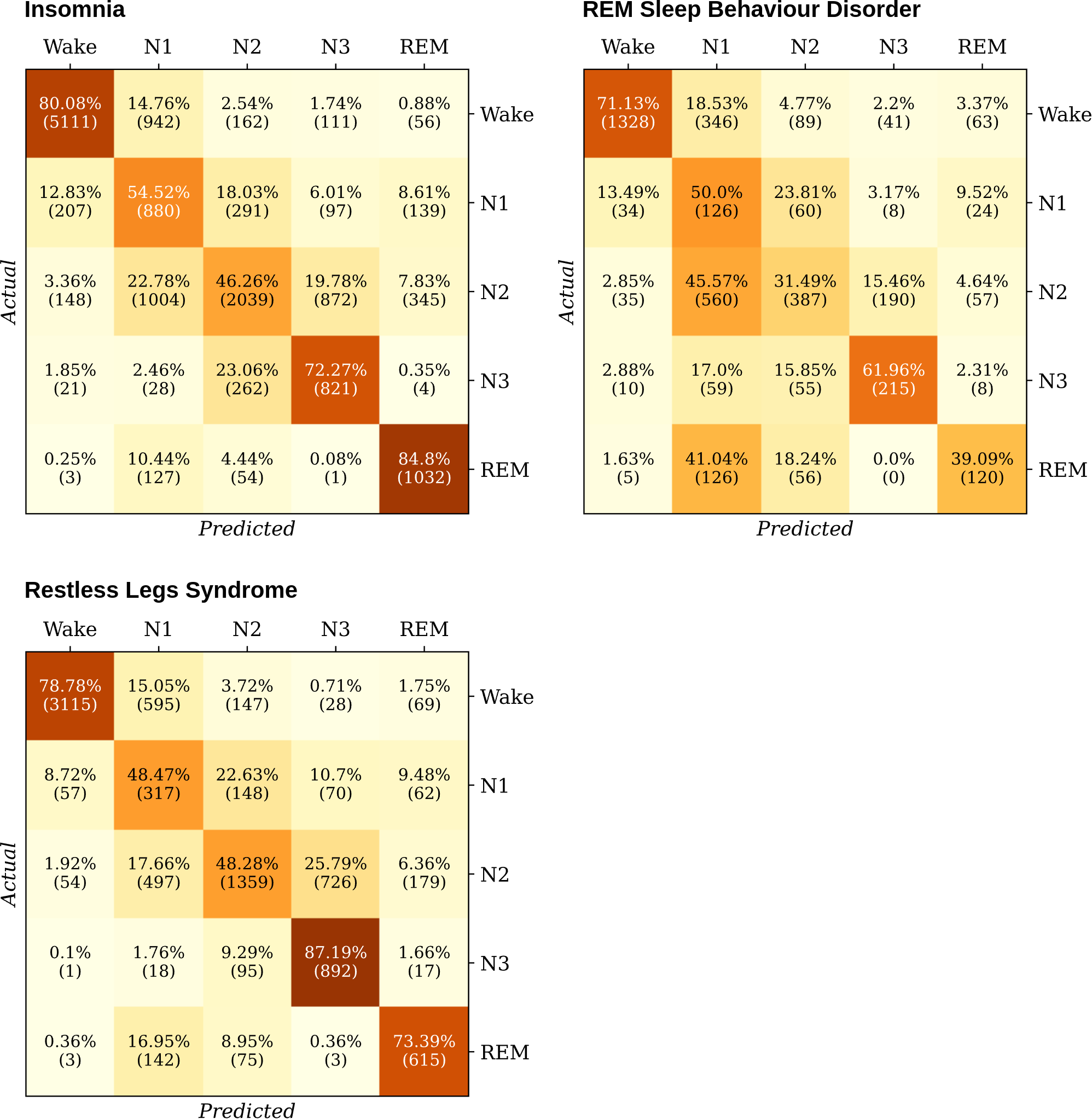}}
  \caption{Confusion matrices for 5-class RNN ensemble's model performance on individuals with sleep disorders in the test set.}
  \label{fig:disorders}
\end{figure}



\pagebreak

\subsection{Accelerometry Signals}\label{sec:hypno-spectrogram}
\begin{figure}[H]
\centerline{\includegraphics[width=0.95\textwidth]{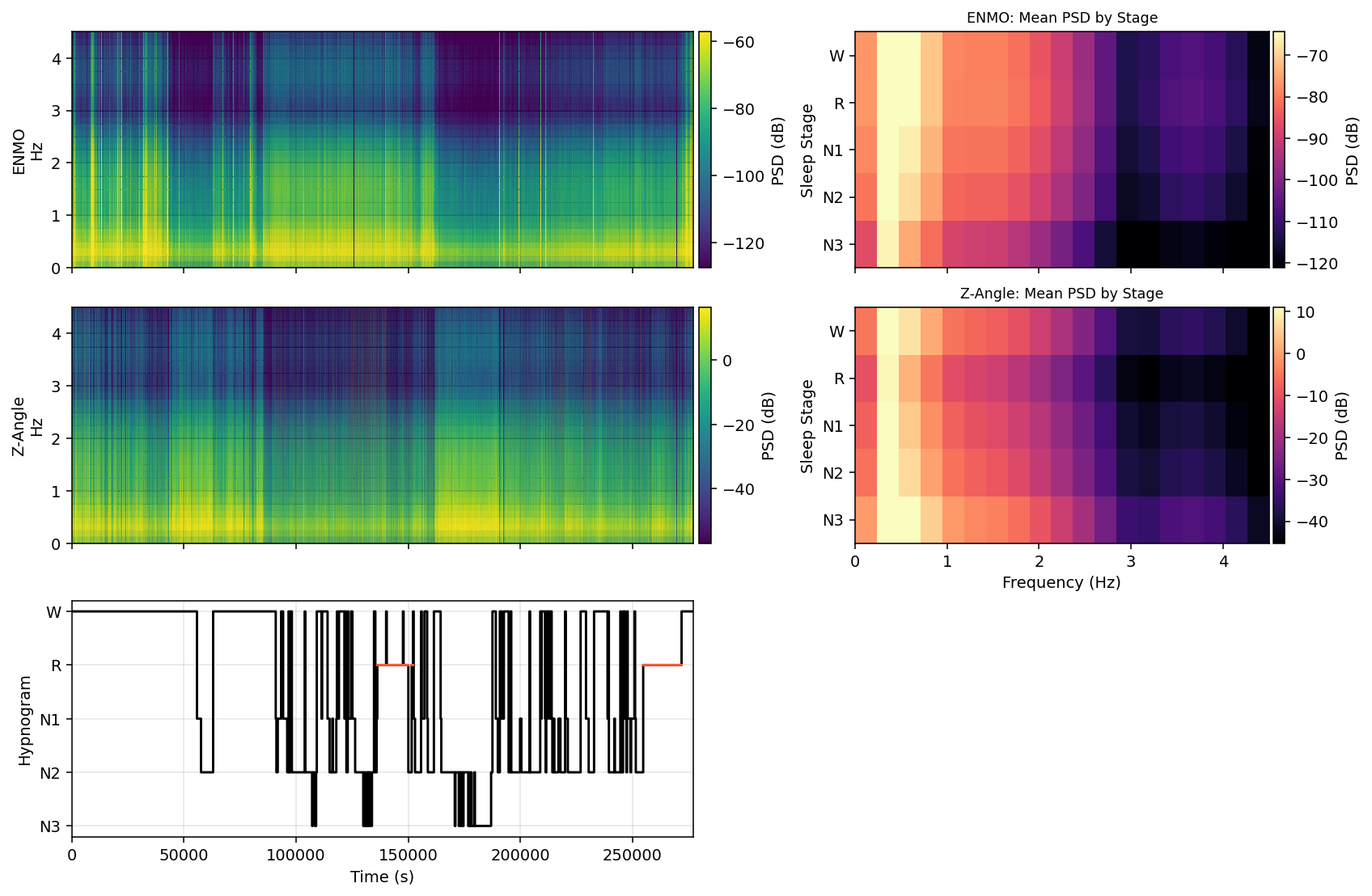}}
\caption{Example 1 of overnight ANNE One chest accelerometer signals (ENMO and z-angle), including spectrograms (left) aligned to the ground-truth hypnogram (bottom), and the mean PSD by sleep stage and frequency across the recording (right).}
  \label{fig:ex1}
\end{figure}

\begin{figure}[H]
\centerline{\includegraphics[width=0.95\textwidth]{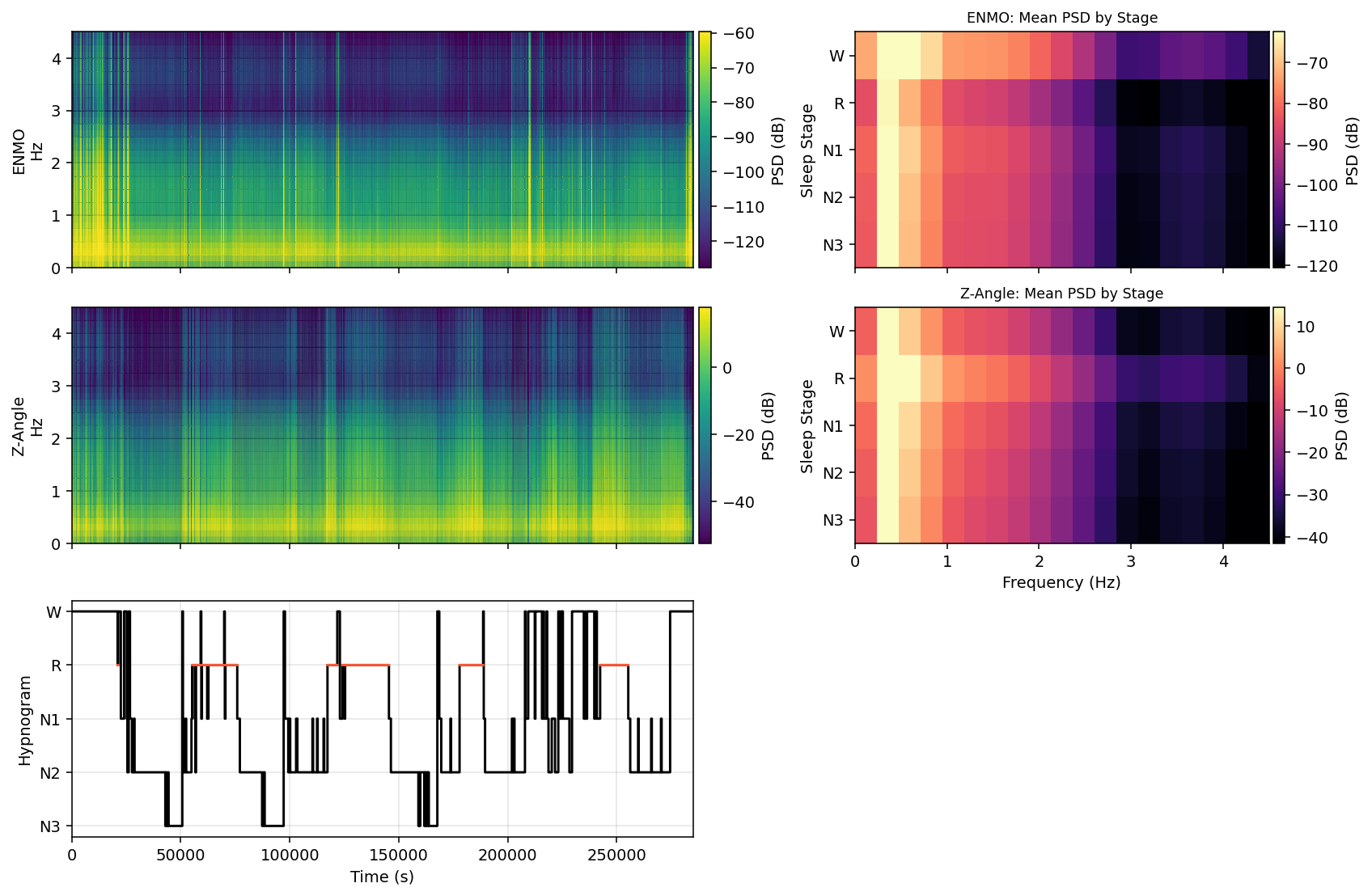}}
\caption{Example 2 of overnight ANNE One chest accelerometer signals (ENMO and z-angle), including spectrograms (left) aligned to the ground-truth hypnogram (bottom), and the mean PSD by sleep stage and frequency across the recording (right).}
  \label{fig:ex2}
\end{figure}

\begin{figure}[H]
\centerline{\includegraphics[width=0.95\textwidth]{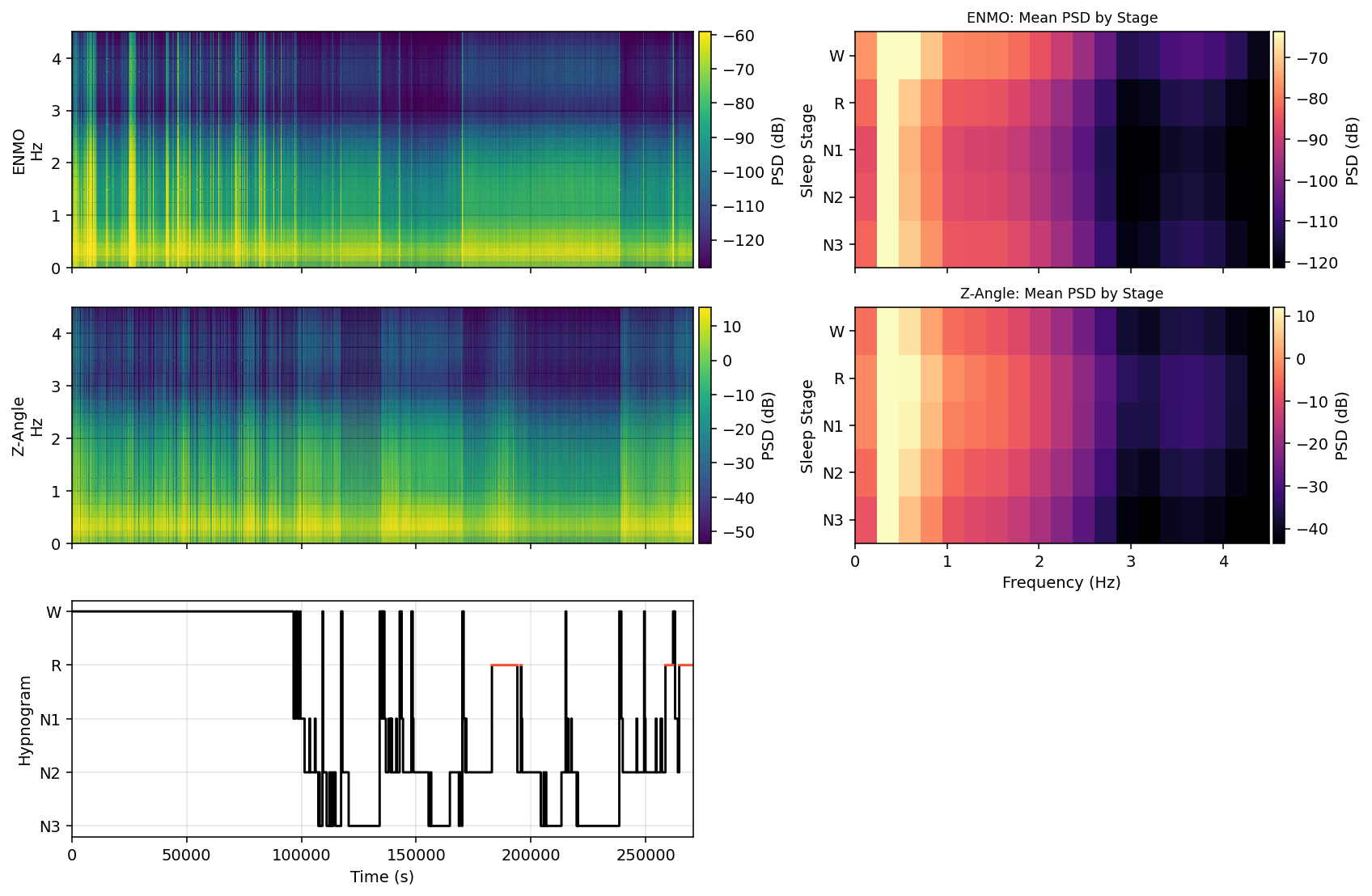}}
\caption{Example 3 of overnight ANNE One chest accelerometer signals (ENMO and z-angle), including spectrograms (left) aligned to the ground-truth hypnogram (bottom), and the mean PSD by sleep stage and frequency across the recording (right).}
  \label{fig:ex3}
\end{figure}

\begin{figure}[H]
\centerline{\includegraphics[width=0.95\textwidth]{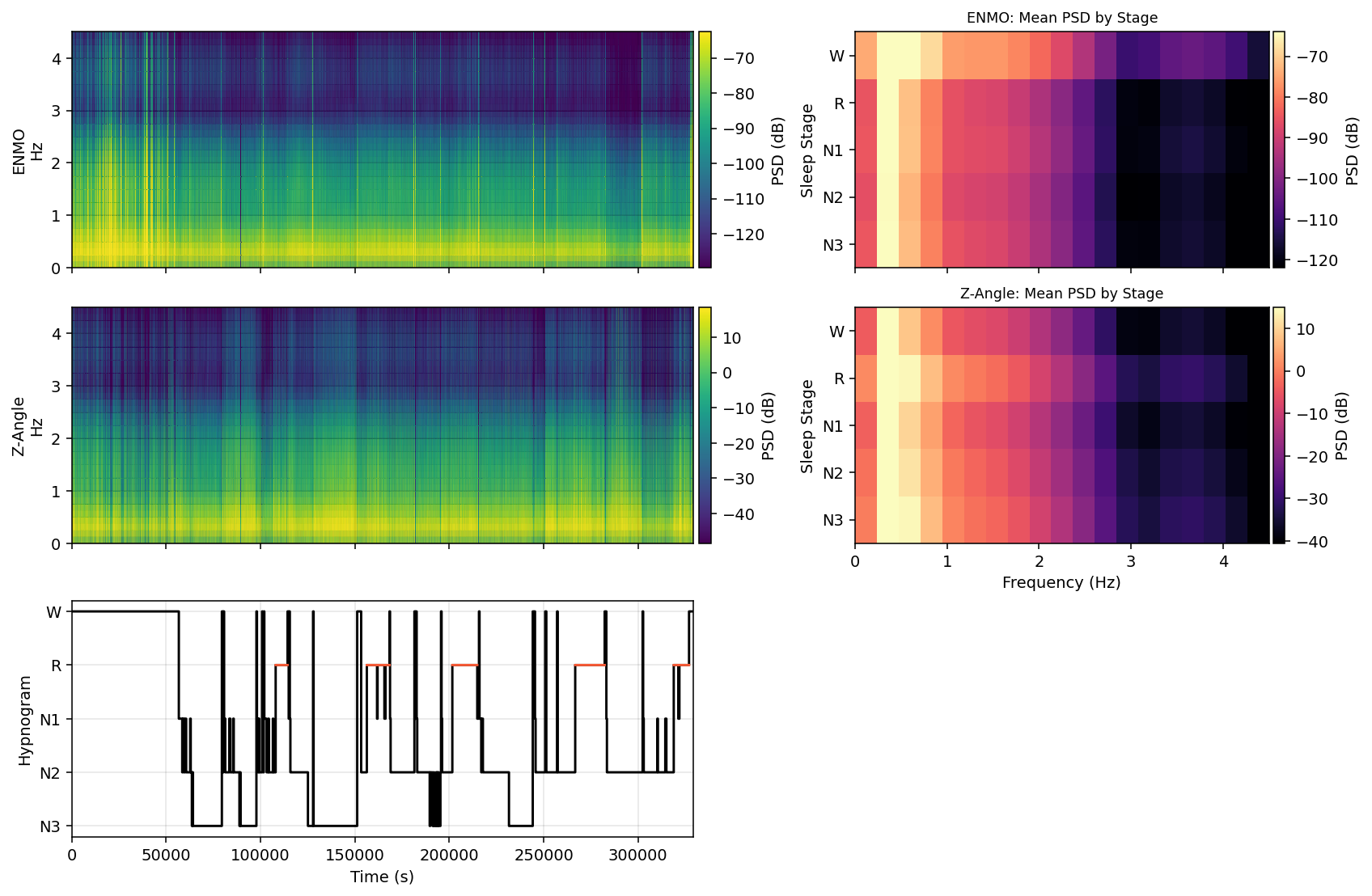}}
\caption{Example 4 of overnight ANNE One chest accelerometer signals (ENMO and z-angle), including spectrograms (left) aligned to the ground-truth hypnogram (bottom), and the mean PSD by sleep stage and frequency across the recording (right).}
  \label{fig:ex4}
\end{figure}

\begin{figure}[H]
\centerline{\includegraphics[width=0.95\textwidth]{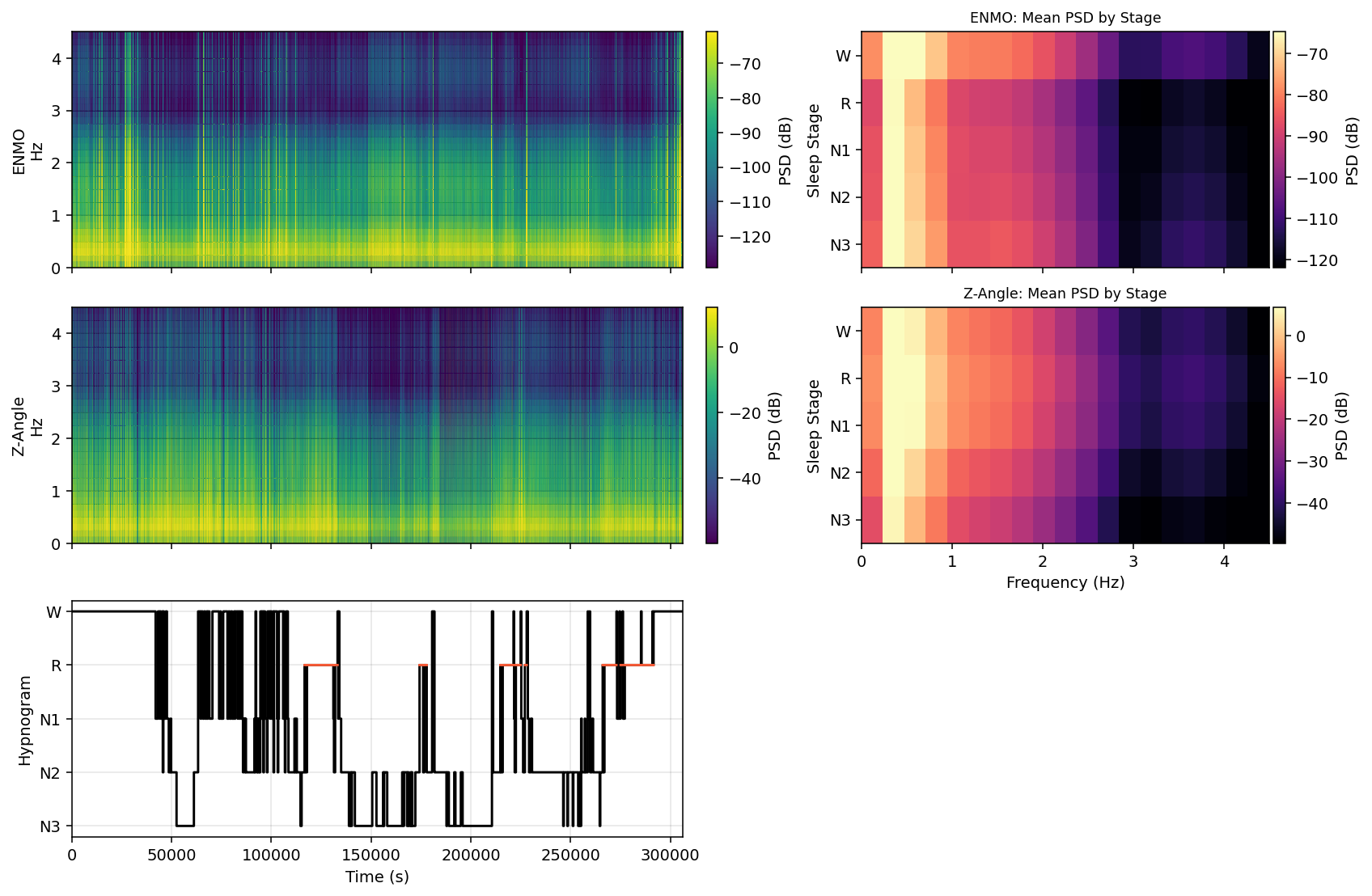}}
\caption{Example 5 of overnight ANNE One chest accelerometer signals (ENMO and z-angle), including spectrograms (left) aligned to the ground-truth hypnogram (bottom), and the mean PSD by sleep stage and frequency across the recording (right).}
  \label{fig:ex5}
\end{figure}

We aggregate our pre-processed accelerometry signals from the ANNE One chest module into Eucliean Norm Minus One (ENMO)$^1$ and z-angle$^2$. On Figures \ref{fig:ex1}, \ref{fig:ex2}, \ref{fig:ex3}, \ref{fig:ex4}, and \ref{fig:ex5}, we select five representative full night recordings of healthy individuals from our study cohort and plot them against the ground truth hypnogram to show how accelerometry informs sleep stage transitions at the individual level. We also show power spectral density (PSD) heatmaps stratified by sleep stage for each recording. 

\def\thefootnote{1}\footnotetext{Euclidean Norm Minus One (ENMO) is calculated as $\sqrt{x^2 + y^2 + z^2} - 1$}

\def\thefootnote{2}\footnotetext{Z-angle is calculated as $\arctan\Big(\frac{z}{\sqrt{x^2 + y^2}}\Big)$}

\vspace{-5mm}

\vspace{-5mm}

\vspace{-5mm}

\pagebreak

\subsection{Accelerometry Signal Ablation Study}\label{sec:accl-ablation}
\begin{table}[H]
\caption{3-class, 4-class, and 5-class evaluation metrics for variants of the No ECG/PPG ensemble model (Accel. + Temp.), including an ablation without chest/limb temperature and a variant with the $0.005-0.2$ Hz band-pass replaced with a $0.5$ Hz high-pass filter.}
\label{tab:accl-ablation}
\footnotesize
\centerline{\begin{tabular}{cr|cccccc}
\Xhline{3\arrayrulewidth} 
\rowcolor[HTML]{EFEFEF} 
\cellcolor[HTML]{EFEFEF} & \cellcolor[HTML]{EFEFEF} & \multicolumn{6}{c}{\cellcolor[HTML]{EFEFEF}\phantom{\Large{A}}\textit{Metric (\%)}\phantom{\Large{A}}} \\
\rowcolor[HTML]{EFEFEF} 
\multirow{-2}{*}{\cellcolor[HTML]{EFEFEF}\textit{Classes}} & \multirow{-2}{*}{\cellcolor[HTML]{EFEFEF}\textit{Feature Set}} & Balanced Accuracy & Precision & Recall & F1 & Cohen's $\kappa$ & MCC \\ \Xhline{3\arrayrulewidth} 

 & \phantom{\Large{A}}Accel. + Temp. & {63.85} & {70.68} & {63.35} & {65.03} & {51.97} & {53.12} \\
& Accel. & 64.60 & 71.15 & 64.18 & 65.64 & 52.97 & 54.12 \\ 
\multirow{-3}{*}{5} 
 & Accel. (0.5 Hz High-Pass)        & 57.40 & 67.00 & 60.84 & 62.31 & 48.21 & 49.12 \\
\hdashline[1pt/1pt]

 & \phantom{\Large{A}}Accel. + Temp. & {73.03} & {74.46} & {71.52} & {72.10} & {58.70} & {59.30} \\
& Accel. & 73.99 & 75.27 & 72.64 & 73.14 & 60.21 & 60.77 \\
\multirow{-3}{*}{4}
 & Accel. (0.5 Hz High-Pass)        & 67.23 & 72.00 & 68.98 & 69.55 & 54.98 & 55.65 \\
\hdashline[1pt/1pt]

 & \phantom{\Large{A}}Accel. + Temp. & {82.78} & {84.26} & {83.18} & {83.51} & {71.71} & {71.56} \\
 & Accel. & 83.08 & 84.82 & 83.76 & 84.08 & 72.47 & 72.61 \\ 
\multirow{-3}{*}{3} 
 & Accel. (0.5 Hz High-Pass)        & 76.11 & 81.48 & 80.13 & 80.60 & 66.45 & 66.64 \\

\Xhline{3\arrayrulewidth} 

\end{tabular}}
\end{table}

\pagebreak

\subsection{Feature Importance by Sensor}
\begin{figure}[H]
\centerline{\includegraphics[width=0.8\textwidth]{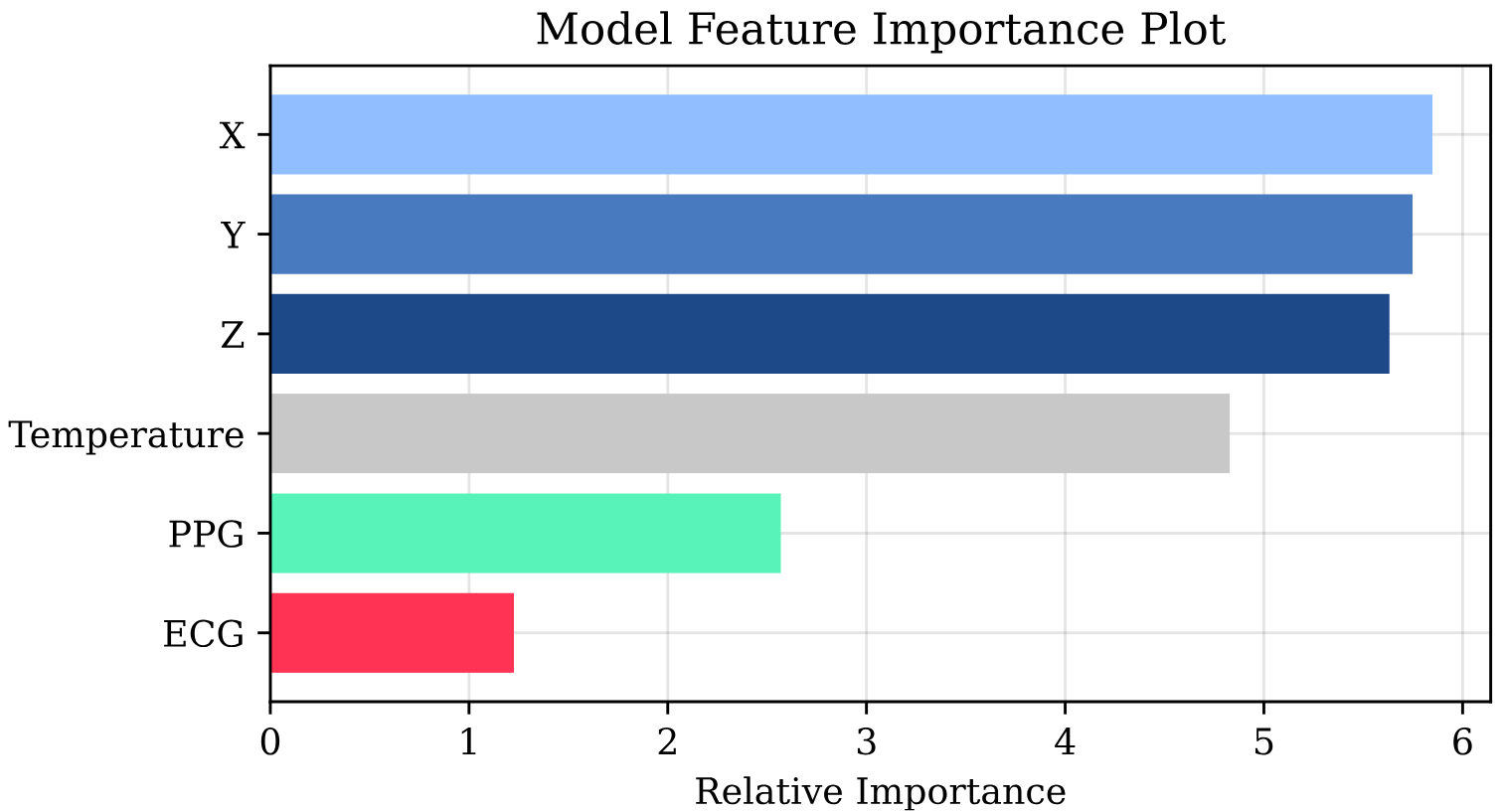}}
\caption{Relative feature importance plot of the 5-class regular model using Integrated Gradients \cite{sundararajan2017axiomatic}, showing the relative importance of the average feature of any sensor compared to another. The values are calculating by taking the absolute value of the contribution followed by averaging across all time windows, then across all recordings, then across features from the same sensor group.}
  \label{fig:feature-importance}
\end{figure}
\end{document}